%% file: arXiv/main.tex
\documentclass[12pt]{article}

\usepackage{paper-macros}

\draftcompilefalse

\includestandalonefalse

\coltsyleproofoffalse

\title{\LARGE \bfseries Protocols for Verifying Smooth Strategies in Bandits and Games}
\author{%
     \begin{minipage}[t]{0.45\textwidth}
         \centering
         {\normalsize Miranda Christ} \\
         {\small \textit{Columbia University}} \\
         {\small$\mathtt{mchrist@cs.columbia.edu}$}
     \end{minipage}
     \hfill
     \begin{minipage}[t]{0.45\textwidth}
         \centering
         {\normalsize Daniel Reichman} \\
         {\small \textit{Worcester Polytechnic Institute}} \\
         {\small$\mathtt{dreichman@wpi.edu}$}
     \end{minipage}
     \vspp \vsp \\
     \begin{minipage}[t]{0.45\textwidth}
         \centering
         {\normalsize Jonathan Shafer} \\
         {\small \textit{MIT}} \\
         {\small$\mathtt{shaferjo@mit.edu}$}
     \end{minipage}
}

\date{~}

\begin{document}

\maketitle

\pagenumbering{gobble} 

\input{parts/abstract}

\newpage

\tableofcontents

\newpage

\pagenumbering{arabic} 

\input{parts/introduction}

\input{parts/preliminaries}

\input{parts/bandits}

\input{parts/games}

\input{parts/conclusion}

\hfil

 \paragraph{Acknowledgments.} MC was supported by a Google CyberNYC grant, an Amazon Research Award, and NSF grants CCF-2312242, CCF-2107187, and CCF-2212233. JS was supported in part by a Simons Investigator Award, an Amazon Research Award, an award from the Research Support Committee at MIT, and NSF grant CNS-2154149.

 \hfil

 \newpage

 \phantomsection

\addcontentsline{toc}{section}{References}
\bibliographystyle{plainnat}
\bibliography{references}

\appendix

\phantomsection

\addcontentsline{toc}{section}{Appendices}

\input{parts/appendix/appendix}

\end{document}

%% file: parts/abstract.tex
\begin{abstract}
We study protocols for verifying approximate optimality of strategies in multi-armed bandits and normal-form games. As the number of actions available to each player is often large, we seek protocols where the number of queries to the utility oracle is \emph{sublinear} in the number of actions. We prove that such verification is possible for sufficiently \emph{smooth} strategies that do not put too much probability mass on any specific action. We provide protocols for verifying that a smooth policy for a multi-armed bandit is $\varepsilon$-optimal. Our verification protocols require provably fewer arm queries than learning. Furthermore, we establish a nearly-tight lower bound on the query complexity of verification in our settings. As an application, we show how to use verification for bandits to achieve verification in normal-form games. This gives a protocol for verifying whether a given strategy profile is an approximate strong smooth Nash equilibrium, with a query complexity that is sublinear in the number of actions.
\end{abstract}

%% file: parts/introduction.tex
\section{Introduction}
An overabundance of available actions makes decision-making difficult in numerous settings, from consumer choice to logistical operations and public policy. 
To gain information about the (often stochastic) reward an action entails, an agent can simply perform that action and experience the outcome. 
However, in many situations it is infeasible for an agent with limited resources to try even a small fraction of the set of available actions. 
One method to mitigate such choice overload is to rely on an external source providing information about the utility of actions. For example, one may consult recommendations from other customers when deciding whether to purchase a product, or solicit advice from consulting firms when formulating a business strategy or a public policy. 
This information, however, may be unreliable, and it is not clear how to efficiently \emph{verify} its accuracy.
Namely, can one outsource information collection to an untrusted external party, and verify correctness by independently assessing the utility of just a few actions?

Verifying utility estimates arises in machine learning (ML) applications as well. 
For example, a client may delegate ML training to a better-resourced but untrusted external party~\citep*{DBLPjournalsjacmGoldwasserKR15,DBLPconfinnovationsGoldwasserRSY21}. 
In \emph{reinforcement learning}, a training algorithm might return a set of estimates for the rewards of possible actions. 
Once again, a challenge for the client is to verify the fidelity of these estimates, which could be inaccurate for numerous reasons. For example, the external party could be malicious, careless and susceptible to errors, or the training algorithm may be faulty.

Similar verification issues emerge in multi-agent settings studied in game theory, where each agent's utility may depend on the other agents' actions.
An object of central interest in such settings is the ($\varepsilon$-approximate) \emph{Nash equilibrium} (NE or $\varepsilon$-NE), which is a strategy profile\footnote{%
    \label{footnote:strategy-profile}%
    A \emph{strategy} is a distribution over actions. A \emph{strategy profile} is an assignment of a strategy to each player (i.e., agent) in the game.%
} 
such that no agent can improve their own utility (by more than $\varepsilon$) by deviating unilaterally to another strategy.
There is a host of questions related to verification and Nash equilibria; specifically, the current paper is motivated by scenarios such as:
\begin{itemize}
    \item{
        \emph{An agent might want to verify that, given that the current strategies of the other agents are known, a proposed strategy is (approximately) the best possible.} For instance, suppose Alice is a corporation navigating a complex market, and she receives advice from a consulting firm that recommends a specific strategy $\tilde{\pi}$. Alice might subsequently like to verify that the proposed $\tilde{\pi}$ is indeed an (approximately) optimal strategy given current market conditions (i.e., given what other market participants are doing).\footnote{%
            This is equivalent to verifying optimality of a proposed strategy for a MAB, as in \cref{section:bandits}. The connection between verification in games and in MAB is developed further in the proofs of our results for games (\cref{section:games}), as outlined in the last paragraph of \cref{section:proof-ideas} below.
        }
    }
    \item{
        \emph{An agent might want to verify that, together with the known current strategy profile of the other agents, a proposed strategy forms an (approximate) Nash equilibrium.} For instance, in the previous example, assume the proposed strategy $\tilde{\pi}$ indeed has optimal expected utility for Alice in current market conditions. Nonetheless, adopting strategy $\tilde{\pi}$ may not be a good idea if, in the case where Alice follows strategy $\tilde{\pi}$, some other market participant, Bob, could unilaterally deviate to a strategy that benefits Bob but significantly harms Alice. Therefore, Alice might like to verify that, together with the current known strategy profile of the other market participants, the proposed strategy $\tilde{\pi}$ forms an (approximate) Nash equilibrium. That would be consistent with $\tilde{\pi}$ being a reasonable strategy for Alice to commit to.
    }
    \item{
        \emph{A mechanism designer might want to verify that a specific mechanism is likely to induce desired behavior.} For instance, suppose Charlie is a governmental agency formulating a new regulation to incentivize market participants to adopt certain behaviors. Even if the regulation itself is easy to understand, it might still be difficult to know how it will influence the market. So Charlie receives advice from a consulting firm, that conducts market research and concludes that, should the new regulation be enacted, market participants are likely to adopt a specific strategy profile $\tilde{\pi}$. Charlie believes that markets gravitate towards Nash equilibria, and therefore, Charlie would like to verify that the strategy profile $\tilde{\pi}$ is indeed an (approximate) Nash equilibrium. If so, that would be consistent with $\tilde{\pi}$ being a likely long-term outcome of the regulation. 
    }

\end{itemize}

We consider two popular models for problems of choosing between different actions: \emph{multi-armed bandits} (MAB) in the single-agent setting, and \emph{normal-form games} in the multi-agent setting.\footnote{For formal defintions of MABs and normal form games please see \cref{definition:bandit,definition:game}.} 
For both MAB and games, we consider settings where $n$, the number of actions available to each player, is a large integer. 
The large number of actions motivates the reliance on advice from external parties. This leads to the following questions:

\begin{question}[Verification of $\varepsilon$-optimal MAB strategy]
    Given a MAB with $n$ arms, is it possible for a verifier to output an $\varepsilon$-optimal strategy, using a number of bandit queries that is sublinear in~$n$ and the help of an untrusted prover?
\end{question}

\begin{question}[Verification of $\varepsilon$-NE]
    Given a $k$-player, $n$-action normal-form game, is it possible for a verifier to output a strategy profile that is an $\varepsilon$-NE, using a number of payoff oracle queries that is sublinear in~$n$ and the help of an untrusted prover?
\end{question}

We study these questions in the framework of \emph{interactive proof systems}~\citep*{DBLPjournalssiamcompGoldwasserMR89}. In this setting, a computationally bounded verifier interacts with an untrusted prover, and can verify the correctness of statements that are intractable to assess without interaction~\citep{DBLPjournalsjacmShamir92}. A recent development in the study of interactive proofs is the application of this methodology to machine learning~\citep{DBLPconfinnovationsGoldwasserRSY21}. We adopt this development in our setting of verification in MABs.\footnote{%
    For formal definitions, see Definitions~\ref{definition:verification-bandits} and \ref{definition:verification-games}.%
}

Let us first give a simple example where the verification of a MAB strategy can be significantly more efficient compared to finding a ``good'' strategy without the help of an untrusted prover. 
Consider the promise problem where a MAB is known to have a single arm with a deterministic reward of $1$, and all other $n-1$ arms have deterministic reward of $0$. It is trivial to \emph{verify} that a given arm communicated by a prover guarantees a reward of $1$ using a single pull to that arm. On the other hand, \emph{finding} such an arm can require $\Omega(n)$ pulls to the arms: if the single arm with reward one is chosen uniformly out of $n$ arms, a linear number of queries to the arms of the bandits is needed to find it~\citep*{DBLPjournalsmlKearnsMN02,DBLPjournalsjmlrMannorT04}.

Similarly, for a given parameter $\gamma$, it can be verified with probability at least $2/3$ that a MAB has an arm with expected reward at least $\gamma-\varepsilon$ for some accuracy parameter $\varepsilon$, using $O(1/\varepsilon^2)$ queries\footnote{We assume throughout this paper that the rewards belong to $[0,1]$} to an arm provided by a prover. As before, finding a strategy with expected reward of $\gamma$ requires $\Omega(n)$ queries to the arms of the MAB.

Observe that in the first example, we have a promise problem in the sense that we know that there exists a single arm with reward $1$ and that this is the arm with the highest reward. In the second example, we verify a \emph{lower bound}
on the value of an optimal strategy, but we do not verify an upper bound on the best possible value, %
hence we do not verify that a given strategy is nearly optimal. There is a good reason for this: even in an interactive proof setting, we show that $\Omega(n/\varepsilon^2)$ queries are necessary (\cref{theorem:linear-lower-bound-for-smooth-bandit-verification}). 
Therefore to achieve verification with a sublinear number of queries to the arms of the bandits, we need to make additional assumptions on either the MAB, or the set of strategies considered. Here we show that verification with a sublinear number of queries to the arms of the bandit is possible for \emph{smooth strategies} as discussed below in \cref{section:smooth-strategies}. 

In interactive proofs, the number of bits communicated between the prover and the verifier may become a bottleneck. Therefore, it is desirable to have \emph{communication efficient} protocols where the number of bits communicated is sublinear in $n$ as well.
We consider communication-efficient protocols in the setting where a prover wishes to convince the verifier of the value of the optimal policy, but does not need to send the optimal policy itself.
We show that such communication-efficient protocols can be achieved for sufficiently smooth MABs by using cryptographic primitives such as vector commitments and SNARKs. %
While our basic protocol consists of just a single message sent from the prover to the verifier, our protocol with a sublinear number of bits of communication uses multiple rounds of interaction. 

The application of cryptography to RL is, to the best of our knowledge, uncommon. We hope that this work can open the door to more such applications.

\section{Verifying smooth strategies: model and results}

\subsection{Smooth strategies: motivation and background}
\label{section:smooth-strategies}

A smooth strategy is a distribution over actions that is not too concentrated on any one action. 
More precisely, a strategy is $\sigma$-smooth if each action is selected with probability at most $\sigma$. Clearly $\sigma \in [1/n,1]$, with smaller values of $\sigma$ corresponding to smoother (i.e., closer to uniform) distributions. To ensure that  our protocols have sublinear query complexity, it is needed that $\sigma n=o(n)$. This is a natural assumption, requiring that the distribution be ``well spread''. 

As a motivation, in many applications each arm in a MAB represents some resource that a system can utilize (e.g., a server in a data center, a driver in a ride-hailing service). Often, as the system grows, the number of available arms increases, as does the total number of arm-pulls, but the capacity of each individual arm remains bounded (e.g., as a ride-hailing service grows, the total number of rides and drivers increase, but the number of rides each individual driver can handle remains the same). If the total number of arm-pulls is, say, linear in the number of arms, then the MAB policy must be $\sigma$-smooth with $\sigma = \BigO{1/n}$.

Additionally, smooth strategies that do not assign too much mass to any particular action can be understood as modeling behavior in various adversarial environments, where players must act in a way that is difficult for an adversary to anticipate. Perhaps surprisingly, our results show that settings with hard-to-predict players allow for efficient (sublinear) verification. 

In a MAB, a strategy is an $\varepsilon$-optimal $\sigma$-smooth strategy if it is $\sigma$-smooth and there is no $\sigma$-smooth strategy with a utility that is greater by more than $\varepsilon$ (Definition~\ref{definition:optimal-smooth-bandit-strategy}).
In a normal-form game, a strategy is an $\varepsilon$-approximate strong $\sigma$-smooth NE if every player's strategy is $\sigma$-smooth, and no player can unilaterally deviate to a $\sigma$-smooth strategy that improves their expected payoff by more than $\varepsilon$.

\cite*{DBLPconfinnovationsDaskalakisGHS24} study smooth strategies in games. They provide additional motivations for studying smooth strategies, and note that a strong $\sigma$-smooth NE (where $\varepsilon=0$) is guaranteed to exist in every normal-form game.

\subsection{Our setting}\label{sub:setting}

\paragraph{Verification in multi-armed bandits.}
In this setting, both a prover and a verifier have access to an $n$-armed bandit.
This access is given via an oracle: one can query the oracle by specifying an arm, and in return receive a reward drawn from that arm's utility distribution.
The prover and verifier communicate interactively, and at the end of the interaction the verifier either rejects or outputs a policy.
If both the prover and the verifier follow the protocol, the verifier should output an approximately optimal smooth policy with probability at least $2/3$; this property is called \emph{completeness}. 
The protocol should also satisfy \emph{soundness}, in that even if the prover behaves arbitrarily, the verifier should output a non-optimal or non-smooth policy with probability at most $1/3$.
Finally, the protocol should be efficient in that both the prover and verifier run in polynomial time, and the verifier makes $o(n)$ queries.\footnote{%
    As usual, it is possible to efficiently amplify the success probability from $2/3$ to $1-\delta$ for $\delta$ arbitrarily small, both in our protocols for bandits and in the protocols for games.%
} For a formal specification, see Definition~\ref{definition:verification-bandits}.

\paragraph{Verification in normal-form games.}
In this setting, both the prover and the verifier have access to a $k$-player, $n$-action normal-form game.
This access is given via a game oracle: one can query the oracle by specifying a strategy profile,\footnote{As in \cref{footnote:strategy-profile}.} and in return receive a vector of payoffs (one per player) corresponding to a tuple of actions drawn from the specified strategy profile.
The prover and the verifier are both given an explicit description of a proposed strategy profile $\pi$, as well as an optimality parameter $\varepsilon$, a smoothness parameter $\sigma$, and a slackness parameter $\eta$.
The prover and verifier communicate interactively, and at the end of the interaction the verifier either accepts or rejects.
Analogously to the MAB setting, this protocol must satisfy completeness and soundness.
Completeness requires that if both parties follow the protocol and $\pi$ is indeed an $\varepsilon$-approximate strong $\sigma$-smooth NE, the verifier accepts with probability at least $2/3$.
Soundness requires that for any (possibly malicious and computationally unbounded) prover, if $\pi$ is not an $(\varepsilon + \eta)$-approximate strong $\sigma$-smooth NE, then the verifier rejects with probability at least $2/3$. For a formal specification, see Definition~\ref{definition:verification-games}.
We note that adapting the MAB verification for verification in games requires certain technicalities such as introducing an additional \emph{slackness parameter} $\eta$, as follows.

\begin{remark}[Slackness]
    In all our verifications settings, there is an approximation parameter $\varepsilon$ and a smoothness parameter $\sigma$. However, in the game verification setting,\footnote{As well as in the variant of bandit verification in \Cref{lemma:verify-smooth-bandit}.} there is an additional slackness parameter $\eta$. This is due to a slight difference between these settings.
    
    When verifying a NE, there is an arbitrary strategy profile $\pi$ provided to the verifier, and we require that the verifier accept \emph{any} input $\pi$ that is an $\varepsilon$-optimal $\sigma$-smooth equilibrium.
    In contrast, in the MAB setting, the verifier outputs an $\varepsilon$-optimal $\sigma$-smooth strategy of its choice; in particular
    in a MAB protocol (e.g., \cref{protocol:bandit-verification}), there may exist $\varepsilon$-optimal $\sigma$-smooth strategies that the verifier never outputs.
    This slight difference necessitates introducing the slackness parameter $\eta$ in the game setting, to create a gap between ``yes'' and ``no'' cases. That is, in verification of NE (Definition~\ref{definition:verification-games}), we require that the verifier accept if $\pi$ is an $\varepsilon$-optimal $\sigma$-smooth NE, and reject if $\pi$ is not an $(\varepsilon+\eta)$-optimal $\sigma$-smooth NE. (If we set $\eta = 0$ in Definition~\ref{definition:verification-games}, then the verifier needs to distinguish whether $\pi$ is $\varepsilon$-optimal, or is $\varepsilon'$-optimal for $\varepsilon' > \varepsilon$ that is arbitrary close to $\varepsilon$. This is not possible when using a stochastic game oracle.)
    Consequently, the MAB verifier's query complexity depends on $\varepsilon$, whereas the NE verifier's query complexity depends on $\eta$.\hfill$\square$
\end{remark}

\subsection{Our contributions}\label{sub:results}

We construct interactive protocols for MABs and normal-form games, where both the verifier and prover have polynomial running time in all parameters.
We also present corresponding lower bounds, showing that verification using our protocols is strictly more efficient than learning in terms of query complexity, and that our protocols have near-optimal dependence on key parameters. In particular:

\begin{enumerate}
\item{
    \textbf{Efficient verification for MABs.} We construct a protocol for verifying $\varepsilon$-optimal $\sigma$-smooth strategies (\cref{theorem:banit-verification}), where the verifier makes $\tildeBigO{\sigma n/\varepsilon^2}$ queries to the MAB oracle. The protocol consists of a single $\tildeBigO{n}$-bit message sent from the prover to the verifier.
}
\item{
    \textbf{Lower bounds for verification of MAB strategies.} We prove a matching lower bound of $\Omega(\sigma n/\varepsilon^2)$ on the number of queries needed by the verifier in \emph{any} such MAB verification protocol (\cref{theorem:linear-lower-bound-for-smooth-bandit-verification}). 
    We also prove that learning approximately optimal smooth strategies requires $\Omega(n)$ queries (\cref{claim:lower-bound-learning-smooth-bandit}), which (together with our verification protocol) implies that verification can be more efficient than learning. For example, when $\sigma = \BigO{1/\sqrt{n}}$, there is a quadratic gap between learning and verifying. 
}
\item{\textbf{Lower communication using cryptography.} We show how to obtain an interactive proof for the value of the optimal smooth policy of a bandit (\Cref{lemma:verify-smooth-bandit-lowcomm}). The asymptotic number of arms pulls is the same as in our original protocol, but the prover now sends at most $\BigO{\secpar \cdot n \sigma \log^3(1/\varepsilon) / \varepsilon}$ bits while affecting the probability of correctness by a negligible additive error: a constant $\lambda$ suffices to ensure an additive error of at most $10^{-6}$. Therefore, assuming $n\sigma/\varepsilon=o(n)$, we have a protocol with sublinear communication.}
\item{
    \textbf{Efficient verification of smooth NE in games.} 
    For normal-form games with $k$ players and $n$ actions, we construct an interactive protocol for verifying that a given strategy profile is an approximate smooth NE, with slackness $\eta > 0$ (\cref{theorem:game-verification}). 
    That is, the verifier accepts if the input strategy profile is an $\varepsilon$-approximate $\sigma$-smooth NE, and it rejects if the input strategy profile is not an $(\varepsilon + \eta)$-approximate $\sigma$-smooth NE.
    The verifier uses $\tildeBigO{k\sigma n/\eta^2}$ queries to the game oracle. 
    Formal statements appear in \cref{section:games}.

    In contrast, for constant $\varepsilon$, Theorem 4 in \cite{DBLPjournalssigecomRubinstein17} states a lower bound of $2^{\OmegaOf{k}}$ queries to the game oracle for computing an $\varepsilon$-NE without the help of a prover, and this lower bound extends also to computing $\sigma$-smooth $\varepsilon$-NE for $\varepsilon$ constant and $\sigma = \ThetaOf{1/n}$ (see Remark 12 in \citealp{DBLPconfinnovationsDaskalakisGHS24}).\footnote{%
        Remark 12 in \citealp{DBLPconfinnovationsDaskalakisGHS24} refers to the case where $\varepsilon$ and $\sigma$ are both constant. However, as discussed in \cref{footnote:deskalakis-smoothness-bandit}, this corresponds to $\sigma = \ThetaOf{1/n}$ in our notation for smoothness.%
    } Thus, our verification protocol offers substantial savings in terms of $k$ as well. We also provide a linear matching lower bound based ideas similar to those of the lower bound for verifying MABs.
}
\item{
    \textbf{Lower bound for verification of smooth NE in games.}
    In \cref{theorem:game-verification-lower-bound}, we show that for verifying a NE, the verifier must make at least $\OmegaOf{kn\sigma}$ queries to the game oracle, matching the performance of our verifier in these parameters up to logarithmic factors. %
}
\end{enumerate}

\subsection{Proof ideas}
\label{section:proof-ideas}

Our bandit verification protocol relies on the following observation. 
Consider a lying prover trying to convince a verifier that a given $\sigma$-smooth bandit strategy $\pi$ is approximately optimal, although there exists another $\sigma$-smooth strategy $\pi^*$ whose expected reward exceeds that of $\pi$ by more than~$\varepsilon$.
Consider a protocol that requires the prover to provide a good estimate of the expected utility of each arm.
Because $\pi^*$ is smooth, it follows that in order to conceal the existence of $\pi^*$, the prover must lie about the utilities of \emph{many} arms.
Thus, it suffices for the verifier to independently estimate the utilities of a few randomly chosen arms, and reject if any of the prover's purported utilities are too far from the verifier's estimates.
Importantly, whereas the prover must lie about \emph{many} arms, it is enough for the verifier to catch \emph{just one} lie.
In particular, if the prover lies about a $\beta$ fraction of utilities, the verifier only needs to query roughly $1/\beta$ arms in order to detect the lie with constant probability. In more detail, assume that the verifier is promised that, if the prover is lying enough to require rejection (the lie distorts the utility of the optimal policy by at least $\varepsilon$), then the prover is in particular lying by more than $\alpha$ on at least a $\beta$ fraction of the arms for some specific known values $\alpha, \beta>0$. This knowledge implies that $\beta n \sigma \alpha \geq \varepsilon$ and also $\alpha \geq \varepsilon$. Based on the promise, the verifier can detect the lie by pulling $O(\frac{1}{\beta}\cdot\frac{1}{\alpha^2})$ arms which by the above inequalities is at most $O(n\sigma/\varepsilon^2).$ 
While $\alpha, \beta$ are not known, the verifier can use a simple partitioning scheme to guess them, which adds a logarithmic term to the query complexity of the verifier.

Our lower bound of $\Omega(\sigma n/\varepsilon^2)$ on the number of queries needed for MAB verification relies on a reduction to the coin bias problem, where one needs to decide with a few samples whether a given coin has bias $1/2-\varepsilon$ or bias $1/2+\varepsilon$. Our proof uses ideas from~\cite*{DBLPconfcoltEvenDarMM02}.

We significantly reduce the communication between the prover and the verifier in our bandit verification protocol using succinct non-interactive arguments of knowledge (SNARKs) and vector commitments (VCs). 
A VC allows one to commit to a vector and reveal individual components whose consistency with the commitment can be proven.
The commitment and opening proofs require space independent of the length of the vector.
A SNARK allows a prover to succinctly prove that a given instance belongs to a polynomial-time computable relation. (See \cref{section:cryptographic-preliminaries} for more details about these cryptographic tools.)

In our NE verification protocol, a central observation is that one can use the bandit verification protocol to ensure that no player has a profitable smooth deviation from a proposed strategy profile $\pi$.
That is, for each $i \in [n]$, if all players except player $i$ behave according to the profile $\pi$, then player $i$ is choosing between $n$ actions, each of which has some fixed reward distribution.\footnote{The reward distribution for each action is fixed, because the reward distribution is a function of the strategies of the remaining players, which are fixed according to $\pi$.} 
This is exactly an $n$-armed bandit.
So player $i$ has no profitable smooth deviation if and only if its current strategy is an optimal smooth strategy for an appropriately defined bandit.
Our NE verification protocol essentially performs this bandit verification $k$ times, once for each player. We also show that a linear dependence on $k$ is necessary.

\section{Related work}

There is a substantial body of work studying algorithms for MABs for finding a distribution over actions (or a single action) that maximizes expected utility~\citep{lattimore2020bandit}. Several works demonstrate a lower bound of $\Omega(n)$ (where $n$ is the number of arms) on the number of arm-pulls that are needed to find a strategy that maximizes utility for the player~\citep*{DBLPconficmlKarninKS13,DBLPjournalscorrChenL15c,DBLPconfcoltChenLQ17,DBLPjournalsjmlrMannorT04,DBLPconfcoltEvenDarMM02,DBLPconfnipsAssadi022}. Specifically, \cite*{DBLPconfcoltEvenDarMM02} provide an algorithm for identifying the best arm up to an additive error of $\varepsilon$ with probability of success at least $1-\delta$. They achieve query complexity of $O(n\log(1/\delta)/\varepsilon^2)$ improving the naive algorithm whose query complexity is $O(n\log(n/\delta)/\varepsilon^2)$. Furthermore, they complement their result by providing a matching lower bound on the query complexity of any algorithm achieving such error guarantees in the PAC framework based on a reduction to the coin bias problem. A lower bound of $\Omega(n\log(1/\delta)/\varepsilon^2)$ was obtained using different methods in~\cite{DBLPjournalsjmlrMannorT04}. They also show a lower bound of $\Omega((\frac{1}{\varepsilon^2})(n+\log(1/\delta)))$ for the case where the set of utilities of the arms is known to the learner, but it does not know which utility belongs to which arm. 
There is also extensive work on regret minimization strategies in MABs~\citep{lai1985asymptotically}, a setting that we do not touch on in this paper. 

A line of research develops learning methods that are robust to adversarial data corruption~\citep*{DBLPconfstocCharikarSV17,DBLPconftccCanettiK21}. The specific case of corrupted rewards (utilities) in multi-armed bandits has received attention as well~\citep*{DBLPconfnipsJun0MZ18,DBLPconfaistatsZhangC0S22}. Their setting differs from ours, as it does not involve an untrusted prover.

A rich line of work examines the complexity of finding Nash equilibria (NE). 
The problem of finding an approximate NE %
in a normal-form game with at least 3 players is known to be complete for the complexity class $\mathsf{PPAD}$ \citep*{DBLPjournalscacmDaskalakisGP09}.
Even for the seemingly simpler case of 2 players, computing an exact NE is $\mathsf{PPAD}$-complete \citep{DBLPconffocsChenD06}, and computing an approximate NE is hard under the Exponential Time Hypothesis for $\mathsf{PPAD}$ \citep*{DBLPjournalssigecomRubinstein17}.
While finding a NE is hard, verifying that a given strategy profile is an (approximate) NE can be done in polynomial time.\footnote{This is true for any problem in $\mathsf{FNP}$, of which $\mathsf{PPAD}$ is a subset.} The study of the query complexity of approximate NE in 2-player normal-form games has received significant attention~\citep*{DBLPjournalssigecomBabichenko19,DBLPjournalsjmlrFearnleyGGS15,DBLPjournalssiamcompGoosR23}. It is known~\citep{DBLPjournalssiamcompGoosR23} that the trivial upper bound of $n^2$ is nearly tight: In certain games $\Omega(n^{2-o(1)})$ queries may be needed to find an approximate NE.

The intractability of finding NE or $\varepsilon$-NE in arbitrary normal-form games motivated \citet*{DBLPconfinnovationsDaskalakisGHS24} to introduce the notion of $\sigma$-smooth NE where each player places a probability mass of at most $1/(n\sigma)$ for each of the $n$ actions and $\sigma$ is a smoothness parameter in $[1/n,1]$ (their parametrization of smoothness differs from ours: recall that we call a strategy smooth if for every action the mass on the action is at most $\sigma$). They consider two equilibrium notions related to smoothness: in a \emph{strong} $\sigma$-smooth NE, each player is executing a smooth strategy and cannot improve their utility by deviating unilaterally to a smooth strategy. In a \emph{weak} $\sigma$-smooth NE, again no player can improve their utility by deviating unilaterally to a smooth strategy; however, the strategies of players in a weak $\sigma$-smooth NE need not be $\sigma$-smooth. Analogous notions are defined by~\citet{DBLPconfinnovationsDaskalakisGHS24} for $\varepsilon$-approximate (smooth) NE where players cannot improve their utilities by more than $\varepsilon$. For strong $\varepsilon$-approximate $\sigma$-smooth equilibrium,~\citet{DBLPconfinnovationsDaskalakisGHS24} prove there exists an algorithm for finding such an equilibrium in time $n^{\BigO{k^4 \log (k/\varepsilon)/\varepsilon^2}}$, where $k$ is the number of players. For weak $\varepsilon$-approximate $\sigma$-smooth NE, they offer an algorithm with runtime complexity independent of $n$ (the runtime depends only on the number of players $k$, the smoothness parameter $\sigma$, and the approximation parameter $\varepsilon$).

Several works study interactive proofs for machine learning~\citep*{DBLPconfinnovationsGoldwasserRSY21,DBLPconfcoltMutrejaS23,DBLPconfstocGurJKRSS24,DBLPconfinnovationsCaroHINS24,DBLPjournalscorrabs241023969}. Their main focus is on verification in supervised learning and similar settings. For instance, verifying that a proposed hypothesis satisfies the agnostic PAC requirement with respect to a (fixed and known) hypothesis class and a (fixed but unknown) population distribution, using less access (samples or queries) to the population distribution than is necessary for agnostic PAC learning. The questions of verifying MAB strategies and NE in normal-form games with few queries to the bandit or game oracle are not studied there.\footnote{However, our definitions of verification for bandits and games are related to the (quite general) Definition 10 in \cite{DBLPconfcoltMutrejaS23} and Definition 4 in \cite{DBLPjournalscorrabs241023969}.}

%% file: parts/preliminaries.tex
\section{Preliminaries}
Let $\bbN = \{1,2,3,\dots\}$. For $n \in \bbN$, we denote by $[n]$ the set $\{1,2,\dots,n\}$ and assume throughout that the set of actions of players both in MABs and games is $[n]$. For a discrete set $\Omega$, we denote by $\distribution{\Omega}$
the set of probability distributions over $\Omega$. We often identify a distribution $p \in \distribution{[n]}$ with the vector $p = (p_1,\dots,p_n)$ such that $p_i = p(i) = \PPP{x \sim p}{x = i}$. Finally, we define the notion of a smooth distribution:

\begin{definition}
    Let $n \in \bbN$ and $\sigma \in [1/n,1]$. A probability distribution $p \in \distribution{[n]}$ is called \ul{$\sigma$-\emph{smooth}} if for every $i \in [n]$, $p_i \leq \sigma$.
\end{definition}

The degree of smoothness of a distribution is governed by the parameter $\sigma$. For $\sigma=1$, smoothness is vacuous as every probability distribution is $1$-smooth. On the other extreme when $\sigma=1/n$ the distribution is the smoothest possible: the uniform distribution.

\subsection{Cryptographic preliminaries}
\label{section:cryptographic-preliminaries}

Let $\secpar \in \bbN$ denote the security parameter.
We write p.p.t.\ to mean probabilistic polynomial time.
We let $\negl$ denote a function that is $O(1/\secpar^c)$ for all $c > 0$.

\subsubsection{Vector commitments}
A \emph{vector commitment} \citep{DBLPconfpkcCatalanoF13} is a tuple of p.p.t.\ algorithms:
\begin{itemize}
	\item $\KeyGen(1^\secpar, n) \to \pp$: takes as input the security parameter and size $n$ of the vectors to be committed, and outputs public parameters $\pp$.
	\item $\Commit_\pp(v) \to c_v, \aux$: takes as input a length-$n$ vector $v$, and outputs a commitment $c_v$ and auxiliary information $\aux$. $\aux$ often contains the entire committed vector $v$.
	\item $\Open_\pp(v_i, i, \aux) \to \pf$: takes as input a value $v_i$, an index $i$, and auxiliary information  $\aux$. It outputs a proof $\pf$ that $v_i$ is the $i^\text{th}$ component of $v$ corresponding to $\aux$.
	\item $\Verify_\pp(c_v, v_i, i, \pf) \to \{\accept, \reject\}$: takes as input a commitment $c_v$, a value $v_i$, an index $i$, and a proof $\pf$. It accepts if and only if $c_v$ commits to a vector whose $i^\text{th}$ component is $v_i$ (except for events with negligible probability).
\end{itemize}

Vector commitments must satisfy \emph{correctness} and \emph{position binding}. 
Correctness requires that with overwhelming probability, any honestly generated public parameters and honestly committed vectors yield valid opening proofs for all of their components.
Position binding requires that it is infeasible for any non-uniform p.p.t.\ adversary to produce a commitment and two valid proofs for \emph{different} openings of that commitment. 
More precisely, for all $n \in \bbZ^+$ and all p.p.t.\ adversaries $\cA$,

\[
\PPP{\pp \gets \KeyGen(1^\secpar, n)}{
\begin{array}{l|l}
	\accept \gets \Verify_\pp(c, v_i, i, \pf) & (c, i, v_i, v_i', \pf, \pf') \gets \cA(1^\secpar, n)\\
	\land\ \accept \gets \Verify_\pp(c, v_i', i, \pf') & 
\end{array}
}
\leq \negl.
\]

We refer the reader to \cite{DBLPconfpkcCatalanoF13} for further details.

\subsubsection{Succinct non-interactive arguments of knowledge}
We present a simplified description of \emph{succinct non-interactive arguments of knowledge} (SNARKs) and refer the reader to \cite{DBLPconfeurocryptGroth16} for full details.

A succinct non-interactive argument of knowledge for a relation generator $\cR$ is a tuple of p.p.t.\ algorithms:
\begin{itemize}
	\item $\Setup(1^\secpar, R) \to \pp, \tau$: takes as input the security parameter and a relation $R \in \cR$, and outputs public parameters $\pp$ and a simulation trapdoor $\tau$.
	\item $\Prove(R, \pp, \phi, w) \to \pf$: takes as input a relation $R$, public parameters $\pp$, and a statement-witness pair $(\phi, w) \in R$.	It outputs a proof $\pf$ of this pair's membership in the relation.
	\item $\Verify(R, \pp, \phi, \pf) \to \{\accept, \reject\}$: takes as input a relation $R$, public parameters $\pp$, a statement $\phi$, and a proof $\pf$. It should accept if and only if $\phi$ has a witness $w$ such that $(\phi, w) \in R$.
\end{itemize}

We consider SNARKs that satisfy \emph{perfect completeness} and \emph{computational knowledge soundness}.

Perfect completeness requires that for all $\secpar \in \bbN$, $R \in \cR$, and $(\phi, w) \in R$:
\[
\PPP{(\pp, \tau) \gets \Setup(1^\secpar, R)}{\pf \gets \Prove(R, \pp, \phi, w) \ : \ \accept \gets \Verify(R, \pp, \phi, \pf)} = 1.
\]

Computational knowledge soundness requires that there exists a non-uniform p.p.t.\ extractor that can extract a witness whenever an adversary can compute an accepting proof.
That is, for all non-uniform adversaries $\cA$, there exists a non-uniform p.p.t.\ extractor $\cX_\cA$ such that
\[
\bbP \left[
\begin{array}{l|l}
	(\phi, w) \notin R \text{ and }& (R, z) \gets \cR(1^\secpar)\\
	\Verify(R, \pp, \phi, \pf) \to \accept & (\phi, \tau) \gets \Setup(1^\secpar)\\
	& ((\phi, w), \pf) \gets (\cA || \cX_{\cA})(R, z, \pp)
\end{array}
\right]
\leq \negl,
\]
where $(\cA || \cX_{\cA})$ denotes that the extractor has access to the adversary's internal state and randomness.

%% file: parts/bandits.tex
\section{Bandits}\label{section:bandits}

\subsection{Definitions}

\begin{definition}[Bandit]
 \label{definition:bandit}
    Let $n \in \bbN$. An \ul{$n$-arm bandit} is a vector of $n$ distributions $q = (q_1,\dots,q_n) \in \left(\distribution{[0,1]}\right)^n$. 
    
    A bandit defines a \ul{bandit oracle} such that, given a query $i \in [n]$ (corresponding to ``pulling the $i$-th arm of the bandit''), the oracle returns a utility $x \sim q_i$ sampled independently of all previous oracle queries and responses. 
    
    The \ul{expected utilities vector} of $q$ is a vector $u = \utility{q} \in [0,1]^n$ such that $u_i = \EEE{x \sim q_i}{x}$ for all $i \in [n]$. 
    A \ul{strategy} for an $n$-arm bandit is a distribution $\pi = (\pi_1,\dots,\pi_n) \in \distribution{[n]}$. The \ul{expected utility of $\pi$ with respect to $u$} is $\EEE{i \sim \pi, x \sim q_i}{x} = \sum_{i = 1}^n \pi_i u_i = \pi \cdot u$.
\end{definition}

\begin{definition}[Smooth bandit strategy]
    \label{definition:smooth-bandit-strategy}
    Let $n \in \bbN$ and $\sigma \in [1/n, 1]$. A strategy $\pi \in \distribution{[n]}$ for an $n$-arm bandit is $\sigma$-smooth if $\pi_i \leq \sigma$ for all $i \in [n]$.
\end{definition}

\begin{definition}[Optimal smooth bandit strategy]
    Let $n \in \bbN$, $\varepsilon \geq 0$, $\sigma \in [1/n, 1]$, let $u \in [0,1]^n$ be the expected utilities vector of an $n$-arm bandit, and let $\pi \in \distribution{[n]}$ be a strategy. We say that $\pi$ is \ul{$\varepsilon$-competitive with respect to $\sigma$-smooth policies for $u$}, if for every $\sigma$-smooth strategy $\pi' \in \distribution{[n]}$, 
    \[
        \pi'\cdot u - \pi \cdot u \leq \varepsilon.
    \]
    If in addition $\pi$ is $\sigma$-smooth, then we say that $\pi$ is an \ul{$\varepsilon$-optimal $\sigma$-smooth strategy for $u$}.\footnote{%
        \label{footnote:deskalakis-smoothness-bandit}
        These are special cases of definitions in \cite{DBLPconfinnovationsDaskalakisGHS24} (see \cref{definition:smooth-nash} below). The first definition corresponds to a \ul{weak $\varepsilon$-approximate $\sigma'\!$-smooth Nash equilibrium for $u$}, and the second definition corresponds to a \ul{strong $\varepsilon$-approximate $\sigma'\!$-smooth Nash equilibrium for $u$}, where $\sigma' = 1/(n\sigma)$.
    }\footnote{%
        We neglect specifying the approximation parameter $\varepsilon$ when it is $0$, and speak simply of strategies that are optimal $\sigma$-smooth, or competitive with $\sigma$-smooth strategies.
    }
    \label{definition:optimal-smooth-bandit-strategy}
\end{definition}

\begin{definition}[Verification of optimality for smooth bandit strategies]
    \label{definition:verification-bandits}
    An \ul{interactive proof system for verification of $\varepsilon$-optimal $\sigma$-smooth policies for $n$-arm bandits} is a pair of algorithms $(\verifier,\prover)$ such that for all $n \in\bbN$, and for every $n$-arm bandit $q$ with expected utilities vector $u = \utility{q} \in [0,1]^n$ and bandit oracle $\cO_q$, and for all $\sigma \in [1/n,1]$ and $\varepsilon \in (0,1)$, the following two conditions hold:
    \begin{itemize}
        \item{
            \textnormal{\textbf{Completeness}}. Let the random variable
            \[
                \pi_\subV = \left[
                    \verifier^{\cO_q}(n,\varepsilon, \sigma)
                    ,
                    \prover^{\cO_q}(n,\varepsilon, \sigma)
                \right]
                \in
                \distribution{[n]} \cup \{\reject\}
            \] 
            denote the output of $\verifier$ after interacting with $\prover$, when each of them receives $(n,\varepsilon, \sigma)$ as input and has oracle access to $\cO_q$. Then
            \[
                \PP{
                    \left(
                        \pi_\subV \neq \reject 
                    \right)
                    ~ \land ~
                    \left(
                        \forall ~
                        \sigma\text{\normalfont{-smooth }}\pi' \in \distribution{[n]}: 
                        ~
                        \pi' \cdot u - \pi_\subV \cdot u \leq \varepsilon
                    \right)
                }
                \geq
                \frac{2}{3}.
            \]
        }
        \item{
            \textnormal{\textbf{Soundness}}. For any (possibly malicious and computationally unbounded) prover $P'$ (which in particular may depend on $n$, $\varepsilon$, $\sigma$ and $q$), the verifier's output $\pi_\subV = \left[
                \verifier^{\cO_q}(n,\varepsilon, \sigma)
                ,
                \prover'
            \right]
            \in
            \distribution{[n]} \cup \{\reject\}$ satisfies
            \[
                \PP{
                    \left(
                        \pi_\subV = \reject 
                    \right)
                    ~ \lor ~
                    \left(
                        \forall ~
                        \sigma\text{\normalfont{-smooth }}\pi' \in \distribution{[n]}: 
                        ~
                        \pi' \cdot u - \pi_\subV \cdot u \leq \varepsilon
                    \right)
                }
                \geq
                \frac{2}{3}.
            \]
        }
    \end{itemize}
    In both conditions, the probability is over the randomness of $\cO_q$ and $\verifier$\!, as well as $\prover$ or $\prover'$.
\end{definition}

We also consider a related notion of verification, where the prover convinces the verifier that the optimal $\sigma$-smooth policy has expected reward in some interval $[t - \varepsilon, t + \varepsilon]$.
In this protocol, the verifier \emph{does not necessarily learn} this policy.
This allows for very low communication between the prover and verifier.

\subsection{Protocols for verifying smooth bandit strategies}
\label{section:bandit-verification}

\begin{protocol}[!ht]
    \begin{ShadedBox}
        \textbf{Assumptions:}
        \begin{itemize}
            \item{
                $n \in \bbN$; $\sigma \in [1/n,1]$; $\varepsilon > 0$.
            }
            \item{
                $q = (q_1,\dots,q_n) \in \left(\distribution{[0,1]}\right)^n$ is an $n$-arm bandit.
            }
            \item{
                $\kP = \kPvalue$.
            }
            \item{
                $\numpulls_\subV(b) = \ceil*{128\lnf{12\cdot(\log_4(1/\varepsilon)+2)\cdot\numarms(b)}/(4^b\varepsilon)^2}$.
            }
            \item{
                $\numarms(b) = \ceil*{4^b\varepsilon \cdot4 n \sigma \cdot (\log_4(1/\varepsilon)+2)\cdot \lnf{6}/\varepsilon}$.
            }
        \end{itemize}

        \vspp

        \textsc{Prover}($n$, $\varepsilon$):
        \begin{algorithmic}
            \For $i \in [n]$
                \For $j \in [\kP]$
                    \State \textbf{sample} $r_{i,j} \sim q_i$
                \EndFor
                \State $\tilde{u}_i \gets \frac{1}{k}\sum_{j = 1}^k r_{i,j}$
            \EndFor
            \vsp
            \State $\tilde{u} \gets (\tilde{u}_1,\dots,\tilde{u}_n)$
            \State \textbf{send} $\tilde{u}$ to verifier
        \end{algorithmic}

        \vspp

        \Verifier($n$, $\varepsilon$, $\sigma$):
        \begin{algorithmic}
            \State \textbf{receive} $\tilde{u}$ from prover
            \vsp
            \For $b \in \set*{0,1,2,\dots,\ceil{\log_4(1/\varepsilon)}}$:
                \FixedWidthComment{Iterate over all bins.}{14em}
                
                \vsp

                \State $\varepsilon_b \gets \varepsilon\cdot4^b$
                \FixedWidthComment{Bin $b$ contains lies of magnitude $| \tilde{u}_i - u_i| \in (\varepsilon\cdot 4^{b-1}, \: \varepsilon\cdot 4^b]$.}{14em}
                \State $\numarms_b \gets a(b)$
                \FixedWidthComment{Number of bandit arms to pull.}{14em}
                \State $\numpulls_b \gets \numpulls_\subV(b)$
                \FixedWidthComment{Number of pulls to each bandit arm.}{14em}
                
                \vsp

                \For $t \in [\numarms_b]$:
                    
                    \vsp
                    
                    \State \textbf{sample} $i_{b,t} \sim \uniform{[n]}$
                    \FixedWidthComment{Select a bandit arm at random.}{14em}
                    
                    \vsp

                    \For $j \in [\numpulls_b]$:
                        \State \textbf{sample} $r_{b,t,j} \sim q_{i_{b,t}}$
                        \FixedWidthComment{Pull (query) the bandit arm.}{14em}
                    \EndFor
                    \State $\hat{u}_{i_{b,t}} \gets \frac{1}{\numpulls_b}\sum_{j = 1}^{\numpulls_b} r_{b,t,j}$
                    \FixedWidthComment{Estimate the bandit arm's utility.}{14em}
                    
                    \vsp

                    \If $|\tilde{u}_{i_{b,t}} - \hat{u}_{i_{b,t}}| > \varepsilon_b/8$:
                        \State \textbf{reject} and terminate execution
                        \FixedWidthComment{Reject if prover's purported utility is far from estimated utility.}{14em}
                    \EndIf
                \EndFor
            \EndFor
            \vsp
            \State{
                $\pi_\subV \gets \ComputeOptimalSmoothBanditStrategy(n, \sigma, \tilde{u})$
            }
            \FixedWidthComment{Compute optimal strategy for prover's purported utilities using \cref{algorithm:optimal-smooth-bandit-strategy}.}{8.2em}
            \vspace*{-3em}
            \State \textbf{output} $\pi_{\subV}$
            \vspp
            \vspp
        \end{algorithmic}
    \end{ShadedBox}
    \caption{A verification protocol for bandits, satisfying the requirements of \cref{theorem:banit-verification}.}
    \label{protocol:bandit-verification}
\end{protocol}

\begin{theorem}[Verification for bandits]
    \label{theorem:banit-verification}
    Let $n \in \bbN$, let $\sigma \in [1/n,1]$, let $\varepsilon \in (0,1)$. \cref{protocol:bandit-verification} defines an interactive proof system $(\verifier,\prover)$ for verification of $\varepsilon$-optimal $\sigma$-smooth policies for $n$-arm bandits such that:
    \begin{itemize}
        \item{
            The protocol consists of a single message of $\BigO{n\logf{1/\varepsilon}}$ bits sent from $\prover$ to $\verifier$.
        }
        \item{
            $\prover$ performs $\mP = \BigO{n\logf{n/\varepsilon}/\varepsilon^2}$ nonadaptive queries to the bandit oracle and runs in time $\poly{n, 1/\varepsilon}$.
        }
        \item{
            $\verifier$ performs
            \[
                \mV = \BigO{
                    \frac{
                        n\sigma
                    }{
                        \varepsilon^2
                    }
                    \cdot \logf{\frac{n\sigma}{\varepsilon}}\logf{\frac{1}{\varepsilon}}
                }
            \]
            nonadaptive queries to the bandit oracle, and runs in time $\poly{n, 1/\varepsilon}$.
        }
    \end{itemize}
    In particular, if $\sigma = \ThetaOf{1/\sqrt{n}}$ then $\mV = \tildeBigO{\sqrt{n}}$, and if $\sigma = \ThetaOf{1/n}$ then $\mV$ is independent of~$n$.
\end{theorem}

The proof of \cref{theorem:banit-verification} appears in \cref{section:banit-verification-proof}.

\subsubsection{A protocol variant for verifying optimality of a given strategy}

\begin{lemma} \label{lemma:verify-smooth-bandit}
	Let $n \in \bbN$, let $\eta, \varepsilon \in (0,1)$, and let $q$ be an $n$-armed bandit. There exists a protocol (\cref{protocol:verify-bandit-variant}) consisting of a prover $\prover$ and a verifier $\verifier$, both of whom are allowed oracle access to $q$ and given $n, \eta, \varepsilon, \pi$ as input.
	The protocol satisfies the following:
	\begin{itemize}
		\item Completeness: If $\pi$ is an $\varepsilon$-optimal $\sigma$-smooth policy for $q$, the verifier $\verifier$ outputs 1 with probability at least $1-\delta$.
		\item Soundness: If $\pi$ is not an $(\varepsilon + \eta)$-optimal $\sigma$-smooth policy for $q$, the verifier $\verifier$ outputs 0 with probability at least $1-\delta$.
	\end{itemize}
In either case:
\begin{itemize}
\item The protocol consists of $O(n \log(1/\eta) \log(1/\delta))$ bits sent from $\prover$ to $\verifier$.
\item $\prover$ performs $O(n (\log(n/\eta)/\eta^2)\log(1/\delta))$ nonadaptive queries to the bandit oracle and runs in time $\poly{n, 1/\eta, 1/\delta}$.
\item $\verifier$ performs
\[
O\left(\frac{n\sigma}{\eta^2} \cdot \log\left(\frac{n \sigma}{\eta}\right) \log\left( \frac{1}{\eta}\right) \log\left( \frac{1}{\delta}\right) \right)
\] nonadaptive queries to the bandit oracle, and runs in time $\poly{n, 1/\eta, 1/\delta}$.
\end{itemize}
\end{lemma}

\cref{protocol:verify-bandit-variant} and the proof for Lemma~\ref{lemma:verify-smooth-bandit} appear in \cref{appendix:bandit-protocol-variant}.

\subsubsection{A low-communication protocol variant}
We observe that using cryptographic tools, specifically succinct non-interactive arguments of knowledge (SNARKs) and vector commitments (VCs), one can significantly reduce the communication between the prover and the verifier in our bandit verification protocol.
A VC allows one to commit to a vector, and reveal individual components whose consistency with the commitment can be proven.
The commitment and opening proofs require space independent of the length of the vector.
A SNARK allows a prover to succinctly prove that a given instance belongs to a polynomial-time computable relation.

We apply these tools in our protocol by having the prover send a commitment to the vector $\tilde{u}$ of purported rewards, rather than sending $\tilde{u}$ in full.
It uses a SNARK to prove that the optimal smooth policy with respect to $\tilde{u}$ has a claimed value.
The verifier then proceeds exactly as in \cref{protocol:bandit-verification}; but instead of examining $\tilde{u}$ directly at each index, it asks the prover for the value and opening proof.

\begin{lemma} \label{lemma:verify-smooth-bandit-lowcomm}
	Let $\secpar \in \bbN$ be a security parameter, let $n \in \bbN$, let $\varepsilon \in (0,1)$, let $q$ be an $n$-armed bandit, and let $u$ denote be the vector of expected utilities of $q$. There exists a protocol (\cref{protocol:bandit-verification-lowcomm}) as follows. 
	The protocol consists of a trusted setup phase, in which shared parameters are generated by a trusted entity; and an interactive phase between a prover and a verifier. 
	Assuming the security of the underlying SNARK $\Pi$ and vector commitment $\VC$, our protocol satisfies:
	\begin{itemize}
		\item \textbf{Completeness:} If the prover behaves honestly, the verifier outputs a value $t$ that is within $\varepsilon$ of the value of the optimal $\sigma$-smooth policy with probability at least $\frac{2}{3} - \negl$.
		\item \textbf{Soundness:} Even if the p.p.t.\ prover behaves arbitrarily, the probability that the verifier outputs a value $t$ that is \emph{not} within $\varepsilon$ of the optimal value is at most $\frac{1}{3} + \negl$.
	\end{itemize}
The efficiency of the protocol is as follows:
    \begin{itemize}
        \item{
            If $\Pi$ has $O(\lambda)$-sized proofs, and $\VC$ has $O(\lambda)$-sized commitments and opening proofs, the protocol consists of $\BigO{\secpar \cdot (\sigma n \log n \log^2(1/\varepsilon) / \varepsilon}$ bits sent between $\prover$ and $\verifier$.
        }
        \item{
            $\prover$ performs $\mP = \BigO{n\logf{n/\varepsilon}/\varepsilon^2}$ nonadaptive queries to the bandit oracle and runs in time $\poly{n, 1/\varepsilon}$.
        }
        \item{
            $\verifier$ performs
            \[
                \mV = \BigO{
                    \frac{
                        n\sigma
                    }{
                        \varepsilon^2
                    }
                    \cdot \logf{\frac{n\sigma}{\varepsilon}}\logf{\frac{1}{\varepsilon}}
                }
            \]
            nonadaptive queries to the bandit oracle, and runs in time $\poly{n, 1/\varepsilon}$.
        }
    \end{itemize}
\end{lemma}
We remark that there exist SNARKs with $(O(\secpar))$-sized proofs for arithmetic circuit satisfiability, which have knowledge soundness in the generic group model \citep[Theorem 2]{DBLPconfeurocryptGroth16}.
There also exist vector commitments with $O(\secpar)$-sized commitments and opening proofs; for example, the CDH-based scheme of \cite[Theorem 5]{DBLPconfpkcCatalanoF13}.

When $n\sigma$ is a constant, the number of bits sent between $\prover$ and $\verifier$ is only $\tilde{O}(\secpar)$, hiding $\log(1/\varepsilon)$ factors.

The protocol and proof for \Cref{lemma:verify-smooth-bandit-lowcomm} appear in \cref{appendix:low-communication-protocol}.

\subsection{Lower bound for bandit verification}
\label{section:lower-bounds-bandit-verification}

\begin{theorem}~\label{theorem:linear-lower-bound-for-smooth-bandit-verification}
    There exist constants $C,c > 0$ as follows. Let $n \in \bbN$, let $\sigma \in [24/n,1]$, and let $\varepsilon \geq 0$. Assume that $(\verifier,\prover)$ is an interactive proof system for verification of $\varepsilon$-optimal $\sigma$-smooth policies for $n$-arm bandits, as in \cref{definition:verification-bandits}. Then there exists an $n$-armed bandit $q$ such that if $(\verifier,\prover)$ are executed with access to the bandit oracle $\cO_q$, then with probability at least $9/10$, $\verifier$ uses at least 
    \[
        \mV \geq C \cdot \frac{\sigma n}{\varepsilon^2} - c = \OmegaOf{\frac{\sigma n}{\varepsilon^2}}
    \] 
    queries to $\cO_q$.
\end{theorem}
\begin{remark}
    \label{remark:verification-lower-bound}
    The linear lower bound in \cref{theorem:linear-lower-bound-for-smooth-bandit-verification} is stated only for $\sigma \in [24/n,1]$, and not for all $\sigma \in [1/n,1]$. Note that for $\sigma = 1/n$ there exists only a single $\sigma$-smooth policy, and so indeed a linear lower bound cannot hold for the entire interval $[1/n,1]$. 
    However, the constants $24$ and $9/10$ appearing in \cref{theorem:linear-lower-bound-for-smooth-bandit-verification} are somewhat arbitrary, and can be replaced by constants closer to $1$.\hfill$\square$
\end{remark}

The proof for \cref{theorem:linear-lower-bound-for-smooth-bandit-verification} appears in \cref{appendix:lower-bounds-for-bandit-verification}.

\subsection{Lower bound for learning smooth bandit strategies}
\label{section:lower-bounds-bandit-learning}

\begin{claim}[Lower bound for learning smooth bandits]
    \label{claim:lower-bound-learning-smooth-bandit}
    Let $n,m \in \bbN$, $n \geq 3$, let $\sigma \in [5/n,1]$, and let $\varepsilon \in [0, 1/4]$. Let $A$ be a (possibly randomized) algorithm such that for any $n$-armed bandit $q$, $A$ performs $m$ (possibly adaptive) oracle queries to the bandit oracle $\cO_q$, and outputs a $\sigma$-smooth strategy $\pi$ such that with probability at least $2/3$, $\pi$ is an $\varepsilon$-optimal $\sigma$-smooth strategy for $q$. Then $m \geq n/6$.
\end{claim}

The proof for \cref{claim:lower-bound-learning-smooth-bandit} appears in \cref{appendix:bandit-learning-lower-bound}.

\begin{remark}
    As in \cref{remark:verification-lower-bound}, the linear lower bound cannot hold for the entire interval $[1/n,1]$. However, the specific constants of $5/n$ and $1/4$ in the statement are somewhat arbitrary.\hfill$\square$
\end{remark}

%% file: parts/games.tex
\section{Games}\label{section:games}

\subsection{Definitions}

\begin{definition}[Game]
\label{definition:game}
    Let $k,n \in \bbN$. A \ul{normal-form game with $k$ players and $n$ actions} is a vector $u = (u_1,\dots,u_k)$ such that for each $i \in [k]$, $u_i: [n]^k \to [0,1]$ is a \ul{utility function} for player $i$. A \ul{strategy profile} for such a game is a vector $\pi = (\pi_1,\dots,\pi_k)$ where for each $i \in [k]$, $\pi_i = (\pi_{i,1},\dots,\pi_{i,n}) \in \distribution{[n]}$ is a \ul{strategy} for player $i$. 
    
    A game defines a \ul{game oracle} $\cO_u$ (corresponding to ``playing the game'') such that given a strategy profile $\pi$, the oracle samples an action vector $a$ by independently sampling $a_1 \sim \pi_1,\dots,a_k \sim \pi_k$, and returns a vector of realized utilities $u(a) = (u_1(a), \dots, u_k(a))$.
    Thus, $\cO_u(\pi)$ is the distribution of realized utilities under $u$ obtained by sampling $a$ from $\pi$.
    
    Given a game $u$ and strategy profile $\pi$, the \ul{expected utility} for a player $i \in [k]$ is
    \[
        \EEE{a_1 \sim \pi_1,\dots,a_k \sim \pi_k}{u_i(a_1,\dots,a_k)}.
    \]
\end{definition}

\begin{definition}[Smooth game strategy]
    Let $k,n \in \bbN$ and $\sigma \in [1/n, 1]$. In a game with $k$ players and $n$ actions, a strategy $\pi_i = (\pi_{i,1}, \dots,\pi_{i,n})$ for a player $i \in [k]$, is \ul{$\sigma$-smooth} if $\pi_{i,j} \leq \sigma$ for each action $j \in [n]$. A strategy profile is $\sigma$-smooth if each strategy in the profile is $\sigma$-smooth.\footnote{
        \label{footnote:deskalakis-sigma-parametrezation}
        This and the following definition correspond to $\sigma'$\!-smoothness in the terminology of \cite{DBLPconfinnovationsDaskalakisGHS24}, where $\sigma' = 1/(n\sigma)$.
    }
\end{definition}

\begin{definition}[Smooth Nash equilibrium; Definition 4 in \cite{DBLPconfinnovationsDaskalakisGHS24}]
    \label{definition:smooth-nash}
    Let $k,n \in \bbN$, $\sigma \in [1/n, 1]$, $\varepsilon \geq 0$. Let $u$ be a game with $k$ players and $n$ actions. A strategy profile $\pi$ for $u$ is a \ul{weak $\varepsilon$-approximate $\sigma$-smooth Nash equilibrium} if for every player $i \in [k]$ and every $\sigma$-smooth strategy $\pi_i' \in \distribution{[n]}$,
    \[
        \EEE{a_i \sim \pi_i', a_{-i} \sim \pi_{-i}}{u_i(a)} - \EEE{a \sim \pi}{u_i(a)} \leq \varepsilon.
    \]
    If in addition $\pi$ is $\sigma$-smooth, we say that $\pi$ is a \ul{strong $\varepsilon$-approximate $\sigma$-smooth Nash equilibrium}.\footnote{%
        We neglect specifying the approximation parameter $\varepsilon$ when it is $0$, and simply speak of (weak or strong) $\sigma$-smooth Nash equilibria. In this paper we focus on strong smooth equilibria. Hence, all smooth NE in this paper are strong smooth NE, unless explicitly mentioned otherwise.
    }
\end{definition}

\begin{definition}[Verification of smooth Nash equilibrium]
    \label{definition:verification-games}
	An \ul{interactive proof system for verification of $\varepsilon$-approximate $\sigma$-smooth Nash equilibria for $k$-player $n$-action games with error $\eta$} is a pair of algorithms $(\verifier,\prover)$ such that for all $k, n \in \bbN$, for every $k$-player $n$-action game $u$, and for all $\sigma \in [1/n, 1]$ and $\varepsilon, \eta \in (0,1)$, the following two conditions hold:
    \begin{itemize}
        \item{
            \textnormal{\textbf{Completeness}}. 
            Let $\pi$ be any $\varepsilon$-approximate $\sigma$-smooth equilibrium.
            Let the random variable 
            \[
            X_\pi = \left[V^{\cO_u}(k, n, \varepsilon, \sigma, \eta, \pi), P^{\cO_u}(k, n, \varepsilon, \sigma, \eta, \pi) \right] \in \{\accept, \reject\}
            \]
            denote the output of $V$ after interacting with $P$, when each of them receives $(k, n, \varepsilon, \sigma, \eta, \pi)$ as input and has oracle access to $\cO_u$. Then
            \[
            \PP{X_\pi = \accept} \geq \frac{2}{3}.
            \]
        }
        \item{
            \textnormal{\textbf{Soundness}}. 
            For any $\pi'$ that is \emph{not} an $(\varepsilon + \eta)$-approximate $\sigma$-smooth equilibrium, and any (possibly malicious and computationally unbounded) prover $P'$ (which in particular may depend on $k, n, \varepsilon, \sigma, \eta, u$, and $\pi'$), the verifier's output $X_{\pi'} = \left[V^{\cO_u}(k, n, \varepsilon, \sigma, \eta, \pi'), P' \right] \in \{\accept, \reject\}$ satisfies
            \[
            	\PP{X_{\pi'} = \reject} \geq \frac{2}{3}.
            \]
        }
    \end{itemize}
\end{definition}

\subsection{Protocol for verifying smooth Nash equilibria}
\label{section:game-verification}

\begin{protocol}[H]
    \begin{ShadedBox}
        \textbf{Assumptions:}
        \begin{itemize}
            \item{
                $k, n \in \bbN$; $\sigma \in [1/n,1]$; $\varepsilon, \eta \in (0,1)$.
            }
            \item{
                $u = (u_1,\dots,u_k)$ is a normal-form game with $k$ players and $n$ actions, where for each $i \in [k]$, $u_i : [n]^k \to [0,1]$.
            }
            \item{
                $\pi = (\pi_1, \ldots, \pi_k)$ is a strategy profile, where for each $i \in [k]$, $\pi_i \in \distribution{[n]}$.
            }
            \item{
            	For each $i \in [k]$, $\cB(i, u, \pi)$ denotes the $n$-arm bandit $q = (q_1, \ldots, q_n)$ where for each $j \in [n]$, $q_j$ is the distribution of player $i$'s utility given that player $i$ plays action~$j$, and the remaining players play according to $\pi$. Namely, $q_j = \cO_u(\pi^{i,j})_i$ where $\pi^{i,j} \in \paren*{\distribution{[n]}}^k$ and for all $i' \in [k]$ and all $j' \in [n]$,
                \[\pi^{i,j}_{i'}(j') = \begin{cases}
                    \pi_{i'}(j') & i' \neq i \\
                    \indicator{j' = j} & i' = i 
                \end{cases}.
                \]
            }
        \end{itemize}

        \vspp
        
        \textsc{Interactive Phase}:
        \vsp 
        \begin{algorithmic}
        	\State Both the prover and verifier are given as input $k, n, \sigma, \varepsilon, \eta$
            \State $\delta \gets 1/(3k)$
            \For $i \in [k]$:
            	\State $b_i \gets \VerifyBandit(n, \cB(i, u, \pi), \sigma, \pi_i, \varepsilon, \delta, \eta)$
            \EndFor
        \end{algorithmic}

        \vspp
        \Verifier:
        \begin{algorithmic}
        \If $b_i = 0$ for any $i \in [k]$:
        	\State \textbf{reject} and terminate execution
        \EndIf	
        \State \textbf{accept}
        \end{algorithmic}
    \end{ShadedBox}
    \caption{A verification protocol for strong smooth Nash equilibria of $k$-player $n$-action games, satisfying the requirements of \cref{theorem:game-verification}.}
    \label{protocol:game-verification}
\end{protocol}

\begin{theorem}[Verification for smooth Nash equilibrium]
    \label{theorem:game-verification}
    Let $k,n \in \bbN$, let $\sigma \in [1/n,1]$, let ${\varepsilon, \eta \in (0,1)}$. \cref{protocol:game-verification} defines an interactive proof system $(\verifier,\prover)$ for verification of $\varepsilon$-approximate $\sigma$-smooth Nash equilibria for $k$-player $n$-action games with slackness $\eta$ such that:
    \begin{itemize}
        \item{
            The protocol consists of $\poly{k, n, 1/\eta}$ rounds between $\prover$ and $\verifier$, with a total of $\poly{k, n, 1/\eta}$ bits sent. 
        }
        \item{
            $\prover$ performs 
            \[
            	\mP = \BigO{
            	\frac{n \log(n/\eta)}{\eta^2} \cdot k\log k
            	}
            \]
             nonadaptive queries to the bandit oracle and runs in time $\poly{k, n, 1/\eta}$.
        }
        \item{
            $\verifier$ performs
            \[
                \mV = \BigO{
                    \frac{n \sigma}{\eta^2}\log\left(\frac{n \sigma}{\eta} \right)\log\left(\frac{1}{\eta} \right)k \log k
                }
            \]
            nonadaptive queries to the game oracle, and runs in time $\poly{k, n, 1/\eta}$.
        }
    \end{itemize}
    In particular, in terms of the dependence on $n$, if $\sigma = \ThetaOf{1/\sqrt{n}}$ then $\mV = \tildeBigO{k\sqrt{n}}$, and if $\sigma = \ThetaOf{1/n}$ then $\mV$ is independent of $n$.
\end{theorem}

The proof of \cref{theorem:game-verification} appears in \cref{appendix:game-verification-protocol}.

\subsection{Lower bound for smooth Nash verification}
\begin{theorem}[Lower bound for verification of smooth Nash equilibrium]
    \label{theorem:game-verification-lower-bound}
    Let $k,n \in\bbN$, $k,n \geq 2$, $\varepsilon \geq 0$, $\eta \in [0,1]$, and let $\sigma \in [2/n,1]$. Assume that $(\verifier,\prover)$ is an  interactive proof system for verification of $\varepsilon$-approximate $\sigma$-smooth Nash equilibria for $k$-player $n$-action games with slackness $\eta$, as in \cref{definition:verification-games}. 
    Then the verifier must use at least 
    \[
        \mV = \OmegaOf{kn\sigma}
    \]
    queries to the game oracle.
\end{theorem}

The proof of \cref{theorem:game-verification-lower-bound} appears in \cref{appendix:game-verification-lower-bound}.

%% file: parts/conclusion.tex
\section{Conclusion}

We have studied how to efficiently verify near optimality of smooth strategies for multi-armed bandits and normal-form games.
In line with previous results~\citep{DBLPconfinnovationsGoldwasserRSY21}, our findings illustrate that verification can be more efficient than learning. Considering the large number of parameters in current machine learning models, we believe that these findings hint that further study of interactive proofs for machine learning models can lead to interesting algorithms with practical applications. One direction is to study verification for reinforcement learning~\cite{DBLPbookslibSuttonB2018} beyond the bandit setting. Another, is to examine interactive proofs for additional game-theoretic solution concepts other than Nash Equilibrium~\cite{DBLPbookscuNRTV2007}, and for extensive form games.

%% file: parts/appendix/appendix.tex
\input{parts/appendix/bandits-proofs/bandit-proofs}

\input{parts/appendix/games-proofs/games-proofs}

\input{parts/appendix/misc-appendices}

%% file: parts/appendix/bandits-proofs/bandit-proofs.tex
\section{Proofs for bandits}

\input{parts/appendix/bandits-proofs/bandit-verification-protocol}

\input{parts/appendix/bandits-proofs/bandit-verification-variant}

\input{parts/appendix/bandits-proofs/bandit-verification-low-communication}

\input{parts/appendix/bandits-proofs/compute-optimal-smooth-strategy}

\input{parts/appendix/bandits-proofs/bandit-verification-lower-bound}

\input{parts/appendix/bandits-proofs/lower-bound-for-learning}

%% file: parts/appendix/bandits-proofs/bandit-verification-protocol.tex
\subsection{Proof of upper bound for verification of smooth bandit strategies}
\label{section:banit-verification-proof}

In this appendix we prove \cref{theorem:banit-verification}.

\begin{claim}
    \label{claim:banit-verification-soundness}
    Let $q \in \left(\distribution{[0,1]}\right)^n$ be an $n$-arm bandit with expected utilities vector $u \in [0,1]^n$. In the context of \cref{protocol:bandit-verification}, let $\tilde{u} \in [0,1]^n$ be the purported expected utilities vector provided by the prover. Assume that there exists a $\sigma$-smooth policy $\pi = (\pi_1,\dots,\pi_n) \in \distribution{[n]}$ such that 
    \[
        \abs*{
            \pi \cdot u
             - 
            \pi \cdot \tilde{u}
        }
        \geq \varepsilon/2.
    \]
    Then the verifier in \cref{protocol:bandit-verification} rejects with probability at least $2/3$.
\end{claim}

\begin{proofof}{\cref{claim:banit-verification-soundness}}
    For each $i \in [n]$, let $\Delta_i = \abs*{u_i - \tilde{u}_i}$. Let $B = \set*{-1,0,1,2,3,\dots,\ceil*{\log_4(1/\varepsilon)}}$, and let 
    \[
        \set*{I_b: ~ b \in B}
    \]
    be a partition of the indices $[n]$ into `buckets', such that for $b \geq 0$,
    \[
        I_b = \set*{i \in [n]: ~ \Delta_i \in \big(\varepsilon\cdot4^{b-1}, \: \varepsilon\cdot4^{b}\big]},
    \]
    and 
    \[
        I_{-1} = \set*{i \in [n]: ~ \Delta_i \leq \varepsilon/4}.
    \]
    By the assumption, 
    \begin{align*}
        \varepsilon/2 
        &\leq 
        \abs*{
            \pi \cdot u
             - 
            \pi \cdot \tilde{u}
        }
        \\
        &\leq
        \sum_{i \in [n]} \pi_i \cdot \abs*{u_i - \tilde{u}_i}
        \\
        &=
        \sum_{i \in [n]} \pi_i \cdot \Delta_i
        \\
        &=
        \sum_{b \in B} \sum_{i \in I_b} \pi_i \cdot \Delta_i
        \\
        &\leq 
        \sum_{b \in B} \sum_{i \in I_b}  \pi_i \cdot \varepsilon\cdot4^{b}
        \tagexplain{%
            $i \in I_b \implies \Delta_i \in \big(\varepsilon\cdot4^{b-1}, \: \varepsilon\cdot4^{b}\big]$%
        }
        \\
        &\leq 
        \varepsilon/4 + \!\!\! \sum_{b \in B\setminus \set{-1}} \sum_{i \in I_b}  \pi_i \cdot \varepsilon\cdot4^{b}
        \tagexplain{$\sum_{i \in[n]}\pi_i \leq 1$}
        \\
        &\leq 
        \varepsilon/4 + \!\!\! \sum_{b \in B\setminus \set{-1}} |I_b| \cdot \sigma \cdot \varepsilon\cdot4^{b}.
        \tagexplain{$\pi$ is $\sigma$-smooth}
    \end{align*}
    Rearranging and dividing by $(|B|-1)\cdot\sigma\cdot\varepsilon$ gives
    \[
        \frac{1}{|B|-1}\sum_{b \in B\setminus \set{-1}} |I_b| \cdot 2^{b} \geq \frac{1}{4\sigma (|B|-1)}.
    \]
    Because the maximum is greater than the average, there exists $b^* \in B$, $b^* \geq 0$ such that  
    \begin{equation}
        \label{eq:proportion-Ibstar}
        |I_{b^*}| 
        \geq 
        \frac{1}{2^{b^*} \cdot 4\sigma \cdot (|B|-1)} 
        \geq 
        \frac{1}{2^{b^*} \cdot 4 \sigma \cdot (\log_4(1/\varepsilon)+2)}
        \geq 
        \frac{\lnf{6}}{\numarms_{b^*}} \cdot n ,
    \end{equation}
    where $\numarms_{b^*}$ is defined as in \cref{protocol:bandit-verification}.

    In the verifier of \cref{protocol:bandit-verification}, consider the iteration of the outer `for' loop in which $b = b^*$. In that iteration, the verifier pulls $\numarms_{b^*}$ arms chosen independently and uniformly at random from $[n]$. The probability that none of these arms belongs to $I_{b^*}$ is
    \begin{equation}
        \label{eq:probability-pull-Ibstar}
        \paren*{
            1-\frac{|I_{b^*}|}{n}
        }^{\numarms_{b^*}}
        \leq
        \paren*{
            1-\frac{\lnf{6}}{\numarms_{b^*}}
        }^{\numarms_{b^*}}
        \leq
        \frac{1}{6},
    \end{equation}
    where we used \cref{eq:proportion-Ibstar} and the inequality $1+x \leq e^{x}$.

    Assume that the verifier pulls some arm $i^* \in I_{b*}$ in iteration $b^*$ of the outer `for' loop in \cref{protocol:bandit-verification}. Then it pulls that arm $\numpulls_{b^*}$ times, and obtains an average utility for those pulls of $\hat{u}_{i^*}$. Let $u_{i^*}$ and $\tilde{u}_{i^*}$ be the true and purported expected rewards for arm $i^*$.
    \begin{align}
        \label{eq:uhat-utilde-diff-large}
        \abs*{\hat{u}_{i^*} - \tilde{u}_{i^*}}
        &\geq
        \abs*{u_{i^*} - \tilde{u}_{i^*}} - \abs*{u_{i^*} - \hat{u}_{i^*}} 
        \tagexplain{Triangle inequality}
        \nonumber
        \\
        &> 
        \varepsilon_{b^*}/4 - \abs*{u_{i^*} - \hat{u}_{i^*}}.
        \tagexplain{%
            $\varepsilon_{b}\! := \varepsilon \cdot 4^{b}$; $i^* \in I_{b^*} \implies \Delta_{i^*} \in \big(\varepsilon_{b^*-1}, \: \varepsilon_{b^*}\big]$%
        }
    \end{align}
    By \cref{theorem:hoeffding},
    \begin{align}
        \label{eq:u-uhat-diff-small}
        \PP{
            \abs*{u_{i^*} - \hat{u}_{i^*}}
            \geq
            \frac{\varepsilon_{b^*}}{16}
        }
        &\leq
        2 \expf{-2 \cdot \paren*{\frac{\varepsilon_{b^*}}{16}}^2 \cdot \numpulls_{b^*}}
        \nonumber
        \\
        &\leq
        2 \expf{-2 \cdot \paren*{\frac{\varepsilon_{b^*}}{16}}^2 \cdot \frac{128\cdot\lnf{12\cdot(\log_4(1/\varepsilon)+2)\cdot\numarms_{b^*}}}{\varepsilon_{b^*}^2}}
        \nonumber
        \\
        &\leq
        \frac{1}{6 \cdot (\log_4(1/\varepsilon)+2)\cdot\numarms_{b^*}} < \frac{1}{6}.
    \end{align}
    Combining \cref{eq:uhat-utilde-diff-large,eq:u-uhat-diff-small} implies that
    \begin{equation}
        \label{eq:uhat-utilde-diff-large-whp}
        \PP{
            \abs*{\hat{u}_{i^*} - \tilde{u}_{i^*}}
            >
            \frac{\varepsilon_{b^*}}{8}
        }
        > \frac{1}{6}.
    \end{equation}
    Finally, applying a union bound to \cref{eq:probability-pull-Ibstar,eq:uhat-utilde-diff-large-whp} implies that with probability at least $1-1/6-1/6 = 2/3$, the verifier pulls an arm $i^* \in I_{b^*}$, and obtains a measurement $\hat{u}_{i^*}$ such that $\abs*{\hat{u}_{i^*} - \tilde{u}_{i^*}}
    > 
    \varepsilon_{b^*}/8$. Therefore, the verifier rejects with probability at least $2/3$, as desired.
\end{proofof}

\begin{claim}
    \label{claim:optimality-for-close-vectors}
    Let $n \in \bbN$, let $\varepsilon \geq 0$, and let $u,v \in [0,1]^n$ such that $\max_{i \in [n]} \: \abs*{u_i - v_i} \leq \varepsilon/2$. Let $\pi^*$ be an optimal $\sigma$-smooth strategy for $u$. Then $\pi^*$ is an $\varepsilon$-optimal $\sigma$-smooth strategy for $v$.
\end{claim}

\begin{proofof}{\cref{claim:optimality-for-close-vectors}}
    For any $\sigma$-smooth strategy~$\pi'$,
    \begin{align*}
            \pi' v - \pi^* v
        &\leq
            (\pi' v - \pi' u) 
            + 
            (\pi' u - \pi^* u) 
            + 
            (\pi^* u - \pi^* v)
        \\
        &\leq
            (\pi' v - \pi' u) 
            + 
            0
            + 
            (\pi^* u - \pi^* v)
            \tagexplain{$\pi^*$ is $\sigma$-smooth optimal for $u$}
        \\
        &\leq
            \sum_{i \in [n]} |u_i - v_i|\cdot\paren*{
                \pi'_i + \pi^*_i
            }
        \\
        &\leq
            \sum_{i \in [n]} (\varepsilon/2)\cdot\paren*{
                \pi'_i + \pi^*_i
            }
            \tagexplain{$\max_{i \in [n]} \: \abs*{u_i - v_i} \leq \varepsilon/2$}
        \\
        &\leq
            \varepsilon.
        \tagexplain{$\pi',\pi^*$ are distributions}
        \qedhere
    \end{align*}
\end{proofof}

\begin{claim}
    \label{claim:banit-verification-completeness}
    Let $q \in \left(\distribution{[0,1]}\right)^n$ be an $n$-arm bandit. If the prover and verifier of \cref{protocol:bandit-verification} interact with each other, and each of them has access to the bandit oracle for $q$, then with probability at least $2/3$, the verifier does not reject, and it outputs a policy $\pi_{\subV}$ that is an $\varepsilon$-optimal $\sigma$-smooth policy with respect to $q$.
\end{claim}

\begin{proofof}{\cref{claim:banit-verification-completeness}}
    Let $u \in [0,1]^n$ be the expected utilities vector of $q$, and let $\tilde{u} \in [0,1]^n$ be the vector of estimates computed by the honest prover, as in \cref{protocol:bandit-verification}. Then
    \begin{align}
        \label{eq:honest-prover-estimates-precise}
        \PP{
            \exists i \in [n]: ~ \big|
                \tilde{u}_i - u_i
            \big|
            > 
            \frac{\varepsilon}{16}
        }
        &\leq
        \sum_{i \in [n]}\PP{
            \big|
                \tilde{u}_i - u_i
            \big|
            > 
            \frac{\varepsilon}{16}
        }
        \tagexplain{Union bound}
        \nonumber
        \\
        &\leq
            2 n \cdot \expf{-2 \kP \cdot \left(\frac{\varepsilon}{16}\right)^2}
        \tagexplain{\cref{theorem:hoeffding}}
        \nonumber
        \\
        &\leq
            2n\cdot \expf{-2 \cdot \frac{128\cdot\lnf{12n/\varepsilon}}{\varepsilon^2} \cdot \left(\frac{\varepsilon}{16}\right)^2}
        <
        \frac{1}{6}.
        \tagexplain{Choice of $\kP$}
    \end{align}
    Similarly, denoting $B = \set*{0,1,2,\dots,\ceil{\log_4(1/\varepsilon)}}$, the verifier's estimates $\hat{u}$ satisfy
    \begin{flalign}
        \label{eq:verifier-estimates-precise}
        &
        \PP{
            \exists b \in B
            ~
            \exists t \in [\numarms_b]:
            ~
            |\hat{u}_{i_{b,t}} - u_{i_{b,t}}| 
            > 
            \frac{\varepsilon_{b}}{16}
        }
        \nonumber
        &
        \\
        &
        \qquad\leq
        \sum_{b \in B}\numarms_b \cdot \PP{
            |\hat{u}_{i_{b,t}} - u_{i_{b,t}}| 
            > 
            \frac{\varepsilon_{b}}{16}
        }
        \tagexplain{Union bound}
        \nonumber
        &
        \\
        &
        \qquad\leq
        \sum_{b \in B}\numarms_b \cdot 2 \expf{-2 \cdot \numpulls_{b} \cdot \paren*{\frac{\varepsilon_{b}}{16}}^2}
        \tagexplain{\cref{theorem:hoeffding}}
        \nonumber
        &
        \\
        &
        \qquad\leq
        \sum_{b \in B}\numarms_b \cdot 2 \expf{-2 \cdot \frac{128\cdot\lnf{12\cdot(\logf{1/\varepsilon}+2)\cdot\numarms_{b}}}{\varepsilon_{b}^2} \cdot \paren*{\frac{\varepsilon_{b}}{16}}^2}
        \nonumber
        &
        \\
        &
        \qquad\leq
        \sum_{b \in B}\numarms_b \cdot \frac{1}{6 \cdot (\logf{1/\varepsilon}+2)\cdot\numarms_{b}} 
        <
        \frac{1}{6}.
        &
    \end{flalign}
    Therefore,
    \begin{flalign*}
        &
        \PP{
            \exists b \in B
            ~
            \exists t \in [\numarms_b]:
            ~
            |\hat{u}_{i_{b,t}} - \tilde{u}_{i_{b,t}}| 
            > 
            \frac{\varepsilon_{b}}{8}
        }
        &
        \\
        &
        \qquad\leq 
        \PP{
            \exists b \in B
            ~
            \exists t \in [\numarms_b]:
            ~
            |\hat{u}_{i_{b,t}} - u_{i_{b,t}}| 
            > 
            \frac{\varepsilon_{b}}{16}
            ~ 
            \lor 
            ~
            |\tilde{u}_{i_{b,t}} - u_{i_{b,t}}| 
            > 
            \frac{\varepsilon_{b}}{16}
        }
        \tagexplain{Triangle inequality}
        &
        \\
        &
        \qquad\leq
        \frac{1}{6} + \frac{1}{6} = \frac{1}{3}.
        \tagexplainhspace{\cref{eq:verifier-estimates-precise,eq:honest-prover-estimates-precise}, union bound}{-5em}
    \end{flalign*}
    This implies that $\PP{G} \geq 2/3$, where $G$ is the event
    \[
        \set*{
            \forall i \in [n]: 
            ~ 
            \big|
                \tilde{u}_i - u_i
            \big|
            \leq
            \frac{\varepsilon}{16}
        }
        ~
        \bigcap 
        ~
        \set*{
            \forall b \in B
            ~
            \forall t \in [\numarms_b]:
            ~
            |\hat{u}_{i_{b,t}} - \tilde{u}_{i_{b,t}}| 
            \leq
            \frac{\varepsilon_{b}}{8}
        }.
    \]
    When $G$ occurs, the verifier does not reject, and by \cref{claim:compute-optimal-smooth-bandit-strategy}, it outputs a strategy $\pi_{\subV}$ that is an optimal $\sigma$-smooth strategy for a bandit with expected utilities vector $\tilde{u}$.
    
    By \cref{claim:optimality-for-close-vectors} and the assumption that event $G$ occurs, $\pi_{\subV}$ is also an $(\varepsilon/8)$-optimal $\sigma$-smooth strategy for $u$.
    
    We conclude that with probability at least $2/3$, the verifier does not reject, and it outputs a strategy $\pi_\subV$ that is (better than) an $\varepsilon$-optimal $\sigma$-smooth strategy for $u$, as desired.
\end{proofof}

\begin{proofof}{\cref{theorem:banit-verification}}
    The completeness property follows from \cref{claim:banit-verification-completeness}.

    For soundness, let $q \in \left(\distribution{[0,1]}\right)^n$ be an $n$-arm bandit with expected utilities vector $u \in [0,1]^n$. Assume the verifier of \cref{protocol:bandit-verification} has access to the bandit oracle for $q$, and it interacts with some (possibly malicious) prover that sends a vector $\tilde{u} \in [0,1]^n$ of purported expected utilities. Consider two cases.
    \begin{itemize}
        \item{
            \textbf{Case I:} There exists a $\sigma$-smooth strategy $\pi \in \distribution{[n]}$ such that $\abs*{
                    \pi \cdot u
                    - 
                    \pi \cdot \tilde{u}
                }
                \geq \varepsilon/2$.
            Then by \cref{claim:banit-verification-soundness}, the verifier rejects with probability at least $2/3$.
        }
        \item{
            \textbf{Case II:} For every $\sigma$-smooth strategy $\pi \in \distribution{[n]}$, $\abs*{
                    \pi \cdot u
                    - 
                    \pi \cdot \tilde{u}
                }
                < \varepsilon/2$. In this case, either the verifier rejects, or by \cref{claim:compute-optimal-smooth-bandit-strategy}, it outputs a strategy $\pi_{\subV}$ that is an optimal $\sigma$-smooth strategy for $\tilde{u}$. Let $\pi^*$ be an optimal $\sigma$-smooth strategy for $u$. Then
                \begin{align*}
                    \pi^*u
                        - 
                    \pi_{\subV}u
                    &=
                    \underbrace{\pi^*u
                    -
                    \pi^*\tilde{u}}_{< \varepsilon/2}
                    +
                    \pi^*\tilde{u}
                    \underbrace{-
                    \pi_{\subV}u
                    +
                    \pi_{\subV}\tilde{u}}_{< \varepsilon/2}
                    -
                    \pi_{\subV}\tilde{u}
                    \\
                    &<
                    \varepsilon
                    +
                    \pi^*\tilde{u}
                    -
                    \pi_{\subV}\tilde{u}
                    \tagexplain{By assumtpion of Case II}
                    \\
                    &\leq \varepsilon,
                    \tagexplain{By optimality of $\pi_{\subV}$ for $\tilde{u}$}
                \end{align*}
                So $\pi_{\subV}$ is an $\varepsilon$-optimal $\sigma$-smooth policy for $u$.
        }
    \end{itemize}
    We conclude that in both cases, with probability at least $2/3$, either the verifier rejects or it outputs an $\varepsilon$-optimal $\sigma$-smooth policy for $u$. This establishes the soundness property.

\end{proofof}

%% file: parts/appendix/bandits-proofs/bandit-verification-variant.tex
\subsection{Protocol variant for verifying optimality of a given strategy}
\label{appendix:bandit-protocol-variant}

In this appendix we prove Lemma~\ref{lemma:verify-smooth-bandit}.

\begin{protocol}[H]
    \begin{ShadedBox}
        \textbf{Assumptions:}
        \begin{itemize}
            \item{
                $n \in \bbN$; $q$ is an $n$-armed bandit.
            }
            \item{
                $\sigma \in [1/n, 1]; \pi \in [0,1]^n; \varepsilon, \eta, \delta \in (0,1)$.
            }
            \item{
                $k = \lceil 18 \ln(8/\delta) \rceil$; $\ell = \lceil \frac{32 \ln(8(k+1)/\delta)}{\eta^2}\rceil$.
            }
        \end{itemize}

        \vspp
        
        \VerifyBandit$(n, q, \sigma, \pi, \varepsilon, \delta, \eta)$:
        \vsp 
        
        \textsc{Interactive Phase}:
        \vsp 
        \begin{algorithmic}
            \For $i \in [k]$:
                \vsp
                \State The prover and verifier run \Cref{protocol:bandit-verification} with parameters $n, \varepsilon = \eta/4, \sigma$.
                \vsp
                \If {the verifier rejects:}
                    \State Let $\pi^{(i)} := \bot$. 
                \Else :
                    \State Let $\pi^{(i)} := $ the strategy output by the verifier.
                \EndIf
            \EndFor
        \end{algorithmic}
            \vsp
            \textsc{Verifier}:
            \begin{algorithmic}
            \vsp 
            \If $\pi$ is not $\sigma$-smooth or not a valid probability distribution:
            \State \textbf{reject} and terminate execution
            \EndIf
            \If $\frac{1}{k}\sum_{i \in [k]} {\indicator{1}}[\pi^{(i)} = \bot] \geq \frac{1}{2}$:
            \State \textbf{reject} and terminate execution
            \EndIf
            \vsp

            \For $i \in [k]$:

                \vsp

                \For $j \in [\ell]$:
                	\State \textbf{if} $\pi^{(i)} \neq \bot$, \textbf{sample} $a \sim \pi^{(i)}$; \textbf{sample} $r_{i,j} \sim q_a$;
                	\State \textbf{else, let}  $r_{i,j} = 0$;
                \EndFor

                \vsp

                $v_i \gets \frac{1}{\ell}\sum_{j \in [\ell]} r_{i,j}$
            \EndFor

            \vsp

            \For $j \in [\ell]$:
                \State \textbf{sample} $a \sim \pi$; \textbf{sample} $r_j \sim q_a$;
            \EndFor

            \vsp

            \State $v \gets \frac{1}{\ell}\sum_{i \in [\ell]} r_j$

            \vsp

            \If $\mathsf{median}\paren*{\{v_i\}_{i \in [k]}} - v > \varepsilon + \eta/2$:
                \State \textbf{reject}
            \EndIf
            \vsp
            \State \textbf{accept}
        \end{algorithmic}
    \end{ShadedBox}
    \caption{A protocol for verifying whether the given bandit strategy is $\eta$-close to an \mbox{$\varepsilon$-approximately} optimal $\sigma$-smooth strategy for the bandit $q$. 
    It uses \Cref{protocol:bandit-verification} as a subroutine.}
    \label{protocol:verify-bandit-variant}
\end{protocol}

\begin{proofof}{Lemma~\ref{lemma:verify-smooth-bandit}}
We first show the following useful fact:

Let $\pi'$ be any valid probability distribution, and let $u$ be the vector of utilities in $[0,1]$ underlying~$q$. If we sample $a \sim \pi'$ and then for each $j \in [\ell]$ we sample $r_j \sim q_a$, then:
\begin{equation}
\PPPunder{\substack{a_j \sim \pi'\\
r_j \sim q_{a_j} \text{ for } j \in [\ell]}}{\left|\pi' \cdot u - \frac{1}{\ell}\sum_{j \in [\ell]} r_j \right| > \eta/8} \leq \frac{\delta}{4(k+1)} \label{eq:estimate-close}    
\end{equation}

This follows from a simple application of \Cref{theorem:hoeffding}, by observing that the expectation of the average of the $r_j$'s is $\pi' \cdot u$:
\begin{align*}
    \PPPunder{\substack{a_j \sim \pi'\\
r_j \sim q_{a_j} \text{ for } j \in [\ell]}}{\left|\pi' \cdot u - \frac{1}{\ell}\sum_{j \in [\ell]} r_j \right| > \eta/8} &\leq 2\exp\left(-2\ell(\eta/8)^2\right)\\
&\leq 2 \exp \left(-\frac{64}{\eta^2} \cdot \ln (8(k+1)/\delta) \cdot \frac{\eta^2}{64} \right)\\
&= 2 \exp \left(-\ln (8(k+1)/\delta) \right)\\
&= \delta/(4(k+1)).
\end{align*}

\paragraph{Completeness.}
We first consider the case where the prover behaves honestly.
Completeness of \Cref{protocol:bandit-verification} ensures that for each protocol run, with probability at least $2/3$ the verifier accepts and the policy sent by the prover is $(\eta/4)$-optimal.
For all $i \in [k]$, define random variables $X_i := 1$ if the verifier accepts, and 0 otherwise.
Since the protocol runs are independent, the $X_i$'s are independent. We can therefore apply \Cref{theorem:hoeffding}:
\begin{align*}
    \PP{\left|\frac{2}{3} - \frac{1}{k} \sum_{i=1}^k X_i\right| > \frac{1}{6}} &\leq 2 \exp\left(-2\ell (1/6)^2 \right)\\
    &\leq 2\exp(-\ln(8/\delta))\\
    &= \delta/4.
\end{align*}
Therefore, strictly more than half of the $\pi^{(i)}$'s will be $(\eta/4)$-optimal.
Furthermore, by \Cref{eq:estimate-close} and a union bound, with probability at least $1 - \delta/2$, all estimates $v_i$ and $v$ will be within $\eta/8$ of the true values of $\pi^{(i)}$ and $\pi$ respectively.
This implies that the median of the $v_i$'s will be within $\eta/8 + \eta/4$ of the optimal value.
Therefore, if the value of $\pi$ is within $\varepsilon$ of optimal, its estimate will be within $\varepsilon + \eta/4 + \eta/8 + \eta/8 = \varepsilon + \eta/2$ of the median of the $v_i$'s and the verifier will accept.

\paragraph{Soundness.} 
We next consider the case where $\pi$ is at least $(\varepsilon + \eta)$-far from optimal, and the prover may behave maliciously.
If at least half of the invocations of \Cref{protocol:bandit-verification} result in rejection, the verifier rejects.
Therefore, for the remainder of the proof we consider the case where more than half of these protocol invocations result in acceptance.

Soundness of \Cref{protocol:bandit-verification} ensures that in each protocol run, the verifier accepts and outputs $\pi$ that is $\eta/4$-far from optimal with probability at most $1/3$.
For all $i \in [k]$, define random variables $X_i := 1$ if $\pi^{(i)}$ is more than $\eta/4$-far from optimal and the verifier accepts; $X_i = 0$ otherwise.
$\PP{X_i = 1}\leq 1/3$; therefore, it follows by the same application of \Cref{theorem:hoeffding} used in Case I that 
\[
\PP{\frac{1}{k}\sum_{i=1}^k X_i \geq 1/2} \leq \delta/4.
\]
Therefore, with probability at least $1 - \delta/4$, strictly more than half of the $\pi_i$'s have value within $\eta/4$ of optimal.
Also with probability at least $1 - \delta/4$, by a union bound, all values $v$ and $v_i$ are $\eta/4$-close to their policies' true values.
If both of these events occur, which happens with probability at least $1 -\delta/2$, the median $v_i$ is at least $\eta/4$-close to optimal.
Since $v$ is within $\eta/4$ of its true value, which more than $(\varepsilon + \eta)$-far from optimal by assumption, $v$ is more than $(\varepsilon + \eta/2)$-far from the median of the $v_i$'s.
Therefore, the verifier will reject.
\end{proofof}

%% file: parts/appendix/bandits-proofs/bandit-verification-low-communication.tex
\subsection{Low-communication protocol for verifying smooth MAB strategies}
\label{appendix:low-communication-protocol}

We use a SNARK, $\Pi$, for the family of relations parameterized by $\VC.\pp$ and $n$:
\[
\cR_{\VC.\pp, n} := 
\left\{
\begin{array}{l|r} 
(c_v, t; v): &\forall i \in [n], v_i \in [0,1]\\
	&\pi \text{ is an optimal } \sigma \text{-smooth policy for }v\\	
	&\pi \cdot v = t\\
	&c_v, \aux_v = \VC.\Commit_{\VC.\pp}(v)
\end{array}
\right\} 
\]

\begin{proof}[Proof of \Cref{lemma:verify-smooth-bandit-lowcomm}]~
    \begin{itemize}
        \item{
            \textbf{Communication.} 
            The prover sends a commitment and SNARK proof, each of which consists of $O(\secpar)$ bits.
            In the interactive phase, the verifier sends $O(a_b  \log(1/\varepsilon)) = O(n \sigma \log^2(1/\varepsilon)/\varepsilon)$ indices, each of which can be written in $\log(n)$ bits.
            The prover sends $O(a_b \log(1/\varepsilon))$ openings and opening proofs, requiring $O(\sigma n \log n \log^2(1/\varepsilon)/\varepsilon)$ bits in total.\footnote{%
                The numbers of queries made by the prover and verifier to the bandit oracle are the same as in \cref{protocol:bandit-verification}.
            }
        }
    \end{itemize}

To show completeness and soundness, we invoke properties of the SNARK and VC to reduce the analysis to that of \cref{protocol:bandit-verification}.

\begin{itemize}
    \item{
        \textbf{Soundness.} 
        Towards a contradiction, consider a p.p.t.\ adversary $\cA$ that acts as the prover and with probability at least $1/3 + 1/\mathsf{poly}(\secpar)$ causes the verifier to output $t$ that is $\varepsilon$-far from the true value of the optimal $\sigma$-smooth policy for some bandit $q$.
        Recall that computational knowledge soundness of $\Pi$ implies that there exists a p.p.t.\ extractor $\cX_\cA$ that, with overwhelming probability, computes $\tilde{u}$ such that $(c_v, t; \tilde{u}) \in \cR_{\VC.\pp, n}$.
        That is, $c_v$ is a commitment to $\tilde{u}$, and the value of the optimal $\sigma$-smooth policy of $\tilde{u}$ is indeed $t$. 
        Now, position binding of the vector commitment implies that with overwhelming probability all openings of $\tilde{u}$ that the prover sends to the verifier are either rejected, or indeed match the corresponding component of $\tilde{u}$.

        Soundness of the protocol now follows exactly from the analysis of \cref{protocol:bandit-verification}.
    }
    \item{
    	\textbf{Completeness.}
    	By construction, $\pi$ is an optimal $\sigma$-smooth policy for $\tilde{u}$, and $(c_{\tilde{u}}, \aux_{\tilde{u}})$ is indeed the output of $\VC.\Commit(\tilde{u})$.
    	Therefore, $(c_v, t; \tilde{u}) \in \cR_{\VC.\pp, n}$ and completeness of $\Pi$ ensures that $\pf$ verifies with probability 1.
    	Correctness of $\VC$ ensures that all vector commitment openings verify with probability $1 - \negl$.
    	The remaining checks performed by the verifier are exactly those from \Cref{protocol:bandit-verification}; therefore, if the prover behaves honestly the verifier accepts with probability at least $2/3 - \negl$.\qedhere
    }
\end{itemize}
\end{proof}

\newpage
\thispagestyle{empty}
\vspace*{-5em}
\begin{protocol}[H]
    \begin{ShadedBox}
        \textbf{Assumptions:}
        \begin{itemize}
            \item{
                $n \in \bbN$; $\sigma \in [1/n,1]$; $\varepsilon \geq 0$; $\lambda \in \bbN$.
            }
            \item{
                $q = (q_1,\dots,q_n) \in \left(\distribution{[0,1]}\right)^n$ is an $n$-arm bandit.
            }
            \item{
                $\kP = \kPvalue$; $\numpulls_\subV(b) = \ceil*{128\lnf{12\cdot(\log_4(1/\varepsilon)+2)\cdot\numarms(b)}/(4^b\varepsilon)^2}$.
            }
            \item{
                $\numarms(b) = \ceil*{4^b\varepsilon \cdot4 n \sigma \cdot (\log_4(1/\varepsilon)+2)\cdot \lnf{6}/\varepsilon}$.
            }
        \end{itemize}

        \vspp
        
        \textsc{Trusted Setup}:
        \begin{algorithmic}
        	\State $\VC.\pp \gets \VC.\KeyGen(1^\secpar, n)$.
        	\State Let $R$ be the relation in $\cR$ parameterized by $\VC.\pp$ and $n$.	
        	\State $\Pi.\pp, \tau \gets \Pi.\Setup(1^\secpar, R)$.
        \end{algorithmic}

        \vspp

        \textsc{Prover}($n$, $\varepsilon, \VC.\pp, \Pi.\pp$):
        \begin{algorithmic}
            \For $i \in [n]$
                \For $j \in [\kP]$
                    \State \textbf{sample} $r_{i,j} \sim q_i$
                \EndFor
                \State $\tilde{u_i} \gets \frac{1}{k}\sum_{j = 1}^k r_{i,j}$
            \EndFor
            \vsp
            \State $\tilde{u} \gets (\tilde{u}_1,\dots,\tilde{u}_n)$
            \State{%
                $\pi \gets \ComputeOptimalSmoothBanditStrategy(n, \sigma, \tilde{u})$; $t \gets \pi \cdot \tilde{u}$
            }
        	\State $c_{\tilde{u}}, \aux_{\tilde{u}} \gets \VC.\Commit_{\VC.\pp}(\tilde{u})$; $\pf \gets \Pi.\Prove(R, \Pi.\pp, (c_{\tilde{u}}, t), \tilde{u})$.
            \State \textbf{send} $c_{\tilde{u}}, t, \pf$ to verifier
        \end{algorithmic}

        \vspp

        \Verifier($n$, $\varepsilon, \VC.\pp, \Pi.\pp$):
        \begin{algorithmic}
            \State \textbf{receive} $c_{\tilde{u}}, t, \pf$ from prover
            \vsp
            \If $\reject = \Pi.\Verify(R, \Pi.\pp, (c_{\tilde{u}}, t), \pf)$:
            	\State \textbf{reject} and terminate execution
            \EndIf
        \end{algorithmic}
        
        \vspp

        \textsc{Interactive phase}:
        \begin{algorithmic}
            \For $b \in \set*{0,1,2,\dots,\ceil{\log_4(1/\varepsilon)}}$:
                \FixedWidthComment{Iterate over all bins.}{14em}

                \State $\varepsilon_b \gets \varepsilon\cdot4^b$; $\numarms_b \gets a(b)$; $\numpulls_b \gets \numpulls_\subV(b)$                
                \vsp

                \For $t \in [\numarms_b]$:
                    
                    \vsp
                    
                    \State Verifier \textbf{samples} $i_{b,t} \sim \uniform{[n]}$
                    \FixedWidthComment{Select a bandit arm at random.}{14em}
                    
                    \vsp

                    \For $j \in [\numpulls_b]$:
                        \State Verifier \textbf{samples} $r_{b,t,j} \sim q_{i_{b,t}}$
                        \FixedWidthComment{Pull (query) the bandit arm.}{14em}
                    \EndFor
                    \State $\hat{u}_{i_{b,t}} \gets \frac{1}{\numpulls_b}\sum_{j = 1}^{\numpulls_b} r_{b,t,j}$
                    \FixedWidthComment{Estimate the bandit arm's utility.}{14em}
                    
                    \vsp
					\State Verifier \textbf{sends} $i_{b,t}$
					\State Prover \textbf{sends} $\pf_{\tilde{u}}, \tilde{u}_{i_{b,t}} \gets \VC.\Open_{\VC.\pp}(c_{\tilde{u}}, i_{b,t}, \aux_{\tilde{u}})$
                	\If $\reject = \VC.\Verify_{\VC.\pp}(c_{\tilde{u}}, \tilde{u}_{i_{b,t}}, i_{b,t}, \pf_{\tilde{u}})$:
                    \FixedWidthComment{Check opening proof.}{10em}
                		\State Verifier \textbf{rejects} and terminates execution 
                	\EndIf
                    \If $|\tilde{u}_{i_{b,t}} - \hat{u}_{i_{b,t}}| > \varepsilon_b/8$:
                        \State Verifier \textbf{rejects} and terminate execution
                    \EndIf
                \EndFor
            \EndFor        
			\vsp
            \State Verifier \textbf{outputs} $t$
        \end{algorithmic}
    \end{ShadedBox}
    \caption{A low communication-complexity verification protocol for bandits, satisfying the requirements of \Cref{lemma:verify-smooth-bandit-lowcomm}.}
    \label{protocol:bandit-verification-lowcomm}
\end{protocol}

%% file: parts/appendix/bandits-proofs/compute-optimal-smooth-strategy.tex
\subsection{Computing an optimal smooth policy for a known bandit}

\begin{algorithmFloat}[H]
    \begin{ShadedBox}
        \textbf{Assumptions:}
        \begin{itemize}
            \item{
                $n \in \bbN$; $\sigma \in [1/n,1]$; $u \in [0,1]^n$.
            }
        \end{itemize}

        \vspp

        \ComputeOptimalSmoothBanditStrategy($n$, $\sigma$, $u$):
        \vsp
        \begin{algorithmic}
            \State{%
                $i_1,\dots,i_n \gets \text{\textbf{sort} } \{1,\dots,n\} \text{ such that } u_{i_1} \geq u_{i_2} \geq \dots u_{i_n}$
            }
            \For $j \in [n]$:
                \State $\pi_{i_j} \gets
                \left\{
                \begin{array}{ll}
                   \sigma  & j \leq \lfloor 1/\sigma \rfloor
                   \\
                   1-\sigma\cdot \lfloor 1/\sigma \rfloor & j = \lfloor 1/\sigma \rfloor + 1
                   \\
                   0 & \text{otherwise}
                \end{array}
                \right.$ 
            \EndFor
            \State $\pi \gets (\pi_1,\dots,\pi_n)$
            \State \textbf{output} $\pi$
        \end{algorithmic}
    \end{ShadedBox}
    \caption{A subroutine of \cref{protocol:bandit-verification}. Computes an optimal $\sigma$-smooth strategy for an $n$-arm bandit with a given expected utilities vector $u$.}
    \label{algorithm:optimal-smooth-bandit-strategy}
\end{algorithmFloat}

Observe that the strategy returned by the above algorithm is indeed $\sigma$-smooth since $1-\sigma\cdot \lfloor 1/\sigma \rfloor \leq \sigma.$

\begin{claim}
    \label{claim:compute-optimal-smooth-bandit-strategy}
    Let $n \in \bbN$, let $\sigma \in [1/n,1]$, and let $q$ be an $n$-armed bandit with expected utilities vector $u \in [0,1]^n$. Then executing \ComputeOptimalSmoothBanditStrategy($n$, $\sigma$, $u$) as in \cref{algorithm:optimal-smooth-bandit-strategy} yields an optimal $\sigma$-smooth strategy $\pi$ for $q$.
\end{claim}

\begin{proofof}{\cref{claim:compute-optimal-smooth-bandit-strategy}}
    Let $\Delta_{\sigma,n} = [0,\sigma]^n$ be the set of all $\sigma$-smooth strategies. $\Delta_{\sigma,n}$ is compact, and the expected utility function $f: ~ \Delta_{\sigma,n} \to \bbR$ given by 
    \[
        f(\pi) = \sum_{i = 1}^n \pi_i u_i
    \]
    is continuous. By the extreme value theorem, $f$ attains a maximum in $\Delta_{\sigma,n}$. 
    
    Let $\pi^* = (\pi^*_1, \dots, \pi^*_n)$ be the strategy constructed by the algorithm. Namely, 
    for indices $i_1,i_2,\dots,i_n\allowbreak \in [n]$ such that $u_{i_1} \geq u_{i_2} \geq \dots \geq u_{i_n}$, we have that $\pi^*_{i_j} = \sigma$ for $j \in \left[\left\lfloor 1/\sigma \right\rfloor\right]$, and the remaining weight, if any, is at index $i_j$ for $j = \left\lfloor 1/\sigma \right\rfloor + 1$.

    We argue that $\pi^*$ is a maximum of $f$ in $\Delta_{\sigma,n}$. Indeed, assume for contradiction that there exists $\pi' \in \Delta_{\sigma,n}$ such that $f(\pi') > f(\pi^*)$.

    Let $s \in [n]$ be the smallest index such that $\pi'_{i_s} u_{i_s} \neq \pi^*_{i_s} u_{i_s}$ (such an index exists because $f(\pi') \neq f(\pi^*)$). Then $\pi'_{i_s} \neq \pi^*_{i_s}$. However, by construction of $\pi^*$, it must be that $\pi'_{i_s} < \pi^*_{i_s} \leq \sigma$.

    Let $t \in [n]$ be the largest index such that $\pi'_{i_t} \neq \pi^*_{i_t}$. Note that $t > s$ and $\pi'_{i_t} > \pi^*_{i_t}$ (otherwise, that would be a contradiction to $f(\pi') > f(\pi^*)$). 
    
    By the choice of $i_1,\dots,i_n$, it must be that 
    \begin{equation}
        \label{eq:uis-greater-uit}
        u_{i_s} > u_{i_t},
    \end{equation}
    since otherwise, $u_{i_s} = u_{i_s+1} = \dots = u_{i_t}$, and that would be a contradiction to $f(\pi') > f(\pi^*)$ (because $\pi'$ and $\pi^*$ differ only on indices $i$ such that $i \in \{i_s, i_{s+1},\dots,i_t\}$).

    Now, let $\delta = \min\,\{\sigma - \pi'_{i_s},\pi'_{i_t}\}$. Note that $\delta > 0$ (because $\pi'_{i_s} < \sigma$ and $\pi'_{i_t} > \pi^*_{i_t} \geq 0$). Consider the strategy $\pi^\delta = (\pi^\delta_1,\dots,\pi^\delta_n)$ which is a modification of $\pi'$ at two entries, such that 
    \begin{itemize}
        \item $\pi^\delta_{i_s} = \pi'_{i_s} + \delta \leq \sigma$,
        \item $\pi^\delta_{i_t} = \pi'_{i_t} - \delta$, and
        \item $\pi^\delta_i = \pi'_i$ for all $i \notin \{i_s,i_t\}$.
    \end{itemize}

    Note that $\pi^\delta$ is $\sigma$-smooth, and furthermore, 
    \begin{align*}
        f(\pi^\delta) - f(\pi')
        &=
        \sum_{i \in [n]} u_i \cdot(\pi^\delta_i - \pi'_i)
        \\
        &=
        u_{i_s}\cdot(\pi^\delta_{i_s} - \pi'_{i_s}) + u_{i_t}\cdot(\pi^\delta_{i_t} - \pi'_{i_t})
        \\
        &= u_{i_s} \cdot \delta + u_{i_t}\cdot(-\delta)
        \\
        &= \delta(u_{i_s} - u_{i_t}) > 0.
        \tagexplain{%
            by \cref{eq:uis-greater-uit} and $\delta > 0$%
        }
    \end{align*}
    This is a contradiction to the maximality of $\pi'$.
\end{proofof}

%% file: parts/appendix/bandits-proofs/bandit-verification-lower-bound.tex
\subsection{Proof of lower bound for verification of smooth bandit strategies}\label{appendix:lower-bounds-for-bandit-verification}

The proof of \cref{theorem:linear-lower-bound-for-smooth-bandit-verification} uses ideas from the proof of Lemma 3 in \cite{DBLPconfcoltEvenDarMM02}. Specifically, it uses a reduction to the coin bias problem, which is defined as follows.
\begin{definition}[Coin Bias Problem]
    Let $\varepsilon,\delta \in (0,1)$ and let $m \in \bbN$. An algorithm $A$ solves the \ul{coin bias problem with precision $\varepsilon$, confidence $1-\delta$, and sample complexity $m$} if for every $b \in \set{-1,1}$, 
    \[
        \PPP{X \sim \paren*{\Ber{\frac{1}{2}+b\varepsilon}}^m}{A(X) = b} \geq 1-\delta.
    \] 
    The probability is over the randomness of $A$ and of $X = (X_1,\dots,X_m)$, which is an i.i.d.\ sample of size $m$ from the Bernoulli distribution with parameter $\frac{1}{2}+b\varepsilon$. 
\end{definition}

The following well-known claim gives a sample complexity lower bound for the coin bias problem (see, e.g., Lemma 5.1 of \citealp{DBLPbooksdaglib0025992}; cf.\ Theorem 11.8.3 in \citealp{DBLPbooksdaglib0016881}).

\begin{claim}[Lower Bound for the Coin Bias Problem]
    \label{claim:lower-bound-for-coins}
    Let $\varepsilon,\delta \in (0,1/4)$ and $m \in \bbN$. Consider the following experiment:
    \begin{enumerate}
        \item{
            Sample $b \sim \uniform{\set{-1,1}}$.
        }
        \item{
            Sample $X_1,X_2,\dots,X_m$ i.i.d.\ from $\Ber{\frac{1}{2} + b\varepsilon}$.
        }
    \end{enumerate}
    Let $f: ~ \zo^m \to \set{-1,1}$. If
    \[
        \PP{f(X_1,X_2,\dots,X_m) = b} \geq 1-\delta
    \]
    then
    \[
        m = \OmegaOf{\frac{\logf{1/\delta}}{\varepsilon^2}}.
    \]
\end{claim}

\begin{figure}[H]
    \centering
    \includegraphics[width=0.8\textwidth]{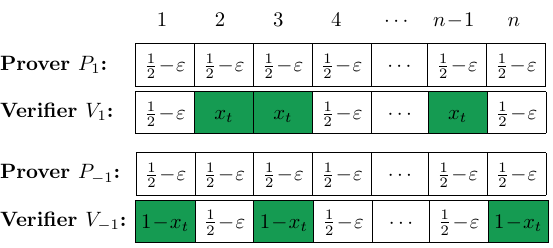}
    \caption{%
        Illustration of the reduction from the coin bias problem to bandit verification. \cref{algorithm:coin-reduction} is given access to a coin with distribution $\Ber{\frac{1}{2}+b^*\cdot\varepsilon}$ for some unknown $b^* \in \set{-1,1}$. \cref{algorithm:coin-reduction} defines two instances of bandit verification, $(\verifier_{1}, \prover_{1})$ and $(\verifier_{-1}, \prover_{-1})$. The prover in both cases has access to a bandit oracle where all arms have utility $\Ber{\frac{1}{2}-\varepsilon}$. The bandit oracles for the verifiers are similar, except that a subset of $1/\sigma$ arms selected at random (depicted here in green) has different utilities. For $\verifier_1$, the reward for these arms is generated by simply flipping the coin each time the arm is queried to get a fresh sample $x_t \sim \Ber{\frac{1}{2}+b^*\cdot\varepsilon}$. For $\verifier_{-1}$ the process is similar, except that the reward is $1-x_t$, i.e., the outcome from the coin is reversed. This design ensures that verifier $\verifier_{-b^*}$ has an oracle where \emph{all} arms have reward $\Ber{\frac{1}{2}-\varepsilon}$, whereas $\verifier_{b^*}$ has a subset of $1/\sigma$ arms with reward $\Ber{\frac{1}{2}+\varepsilon}$.
    }
    \label{figure:coin-to-bandit-reduction}
\end{figure}

The reduction from the coin bias problem to bandit verification is depicted in \cref{figure:coin-to-bandit-reduction}.

\begin{algorithmFloat}[H]
    \begin{ShadedBox}
        \textbf{Assumptions:}
        \begin{itemize}
            \item{
                $n,m \in \bbN$, $\sigma \in [24/n,1]$, $\varepsilon \geq 0$. For simplicity, assume $1/\sigma \in \bbN$.
            }
            \item{
                $(\verifier,\prover)$ is an interactive proof system for verification of $\varepsilon$-optimal $\sigma$-smooth policies for $n$-arm bandits, as in \cref{definition:verification-bandits}.
            }
            \item{
                $x_1,x_2,\dots,x_m \in \zo$ are sampled as in the coin experiment of \cref{claim:lower-bound-for-coins}.
            }
        \end{itemize}

        \vspp

        \textsc{DecideCoinBias}($x_1,x_2,\dots,x_m$):
        \begin{algorithmic}
            \vsp
            \For $b \in \set{-1,1}$:
                \State \textbf{sample} $I_b \sim \uniform{\binom{[n]}{1/\sigma}}$
                \Comment{%
                    $I_b \subseteq [n]$ is a subset of cardinality $|I_b| = 1/\sigma$.%
                }
                \State $(\verifier_b,\prover_b) \gets $ fresh copy of $(\verifier,\prover)$
            \EndFor

            \vsp

            \State $q \gets \big(\!\Ber{1/2-\varepsilon},\dots,\Ber{1/2-\varepsilon}\big)$
            \Comment{$q$ is an $n$-arm bandit.}
            \vsp
            \State $b \gets 1$
            \Comment{Simulate interactive proof $(\prover_b,\verifier_b)$.}
            \vsp
            \State $t \gets 1$
            \vsp
            \While $t \leq m$:
                \Comment{Use coin sample $x_t$ for $t = 1,2,\dots,m$.}
                \vsp

                    \State \textbf{continue simulation} of $(\verifier_b,\prover_b^{\cO_q})$ until $\verifier_b$ terminates or queries an oracle arm
                    \vsp

                    \If $\verifier_b$ queried an arm $i \in [n]$:
                        \vsp
                        \If $i \in I_b$:
                            \vsp
                            \State $r \gets \left\{
                                    \begin{array}{ll}
                                        x_t & \quad b = 1 \\
                                        1-x_t & \quad b = -1
                                    \end{array}
                                    \right.$
                            \Comment{Simulate arm $i \in I_b$ using coin sample $x_t$.}
                            \vsp
                            \State \textbf{send} $r$ to $\verifier_b$
                            \vsp
                            \State $b \gets (-b)$
                            \Comment{Switch to simulating $(\prover_{-b},\verifier_{-b})$.}
                            \vsp
                            \State $t \gets t+1$
                            \Comment{Proceed to next coin sample.}
                            \vsp
                        \Else :
                            \vsp
                            \State \textbf{sample} $r \sim \Ber{1/2 - \varepsilon}$
                            \Comment{Simulate arm $i \notin I_b$.}
                            \vsp
                            \State \textbf{send} $r$ to $\verifier_b$
                        \EndIf
                    \Else :
                    \Comment{$\verifier_b$ has terminated.}
                        \vsp
                        \State $\pi \gets$ output returned by $\verifier_b$
                        \vsp
                        \If $\pi = \reject$ \textbf{or} $\sum_{i \in I_b} \pi_{i} \geq 1/2$:
                            \vsp
                            \hypertarget{line:return-b}{}
                            \State \textbf{output} $b$ and \textbf{terminate}
                            \vsp
                        \Else :
                            \vsp
                            \hypertarget{line:return-minus-b}{}
                            \State \textbf{output} $-b$ and \textbf{terminate}
                        \EndIf
                    \EndIf
            \EndWhile

            \vspp

            \State \textbf{output} $\bot$ and \textbf{terminate}
            \Comment{We have exhausted our sample supply $x_1,\dots,x_m$.}
        \end{algorithmic}
    \end{ShadedBox}
    \caption{A reduction from the coin bias problem of \cref{claim:lower-bound-for-coins} to verification for smooth bandits (as in \cref{definition:verification-bandits}).}
    \label{algorithm:coin-reduction}
\end{algorithmFloat}

\begin{proofof}{\cref{theorem:linear-lower-bound-for-smooth-bandit-verification}}
    For simplicity, assume that $1/\sigma \in \bbN$. We show a reduction from the coin bias problem of \cref{claim:lower-bound-for-coins} to verification for smooth bandits (\cref{definition:verification-bandits}).
    
    The reduction is given in \cref{algorithm:coin-reduction}, which solves the coin bias problem by simulating two copies of an interactive proof $(\verifier,\prover)$ for bandit verification, denoted $(\verifier_1,\prover_1)$ and $(\verifier_{-1},\prover_{-1})$.

    In both copies, the prover is given oracle access to an $n$-armed bandit where all arms have utility $\Ber{1/2-\varepsilon}$. For each $b \in \set{-1,1}$, the verifier $\verifier_b$ is given access to a bandit oracle corresponding to a subset $I_b \subseteq [n]$ of cardinality $1/\sigma$ that is chosen uniformly at random. For each arm $i \notin I_b$, the utility is distributed $\Ber{1/2-\varepsilon}$, the same as in the prover's oracle. However, for arms $i \in I_b$, the utility is simulated using a sequence of i.i.d.\ samples $x_1, x_2, \dots, x_m$ from the coin bias problem. Specifically, for verifier $V_1$, the utility for an arm $i \in I_1$ is simulated by returning $x_t$, where $x_t$ is the next unused sample from the sequence $x$. In contrast, for verifier $V_{-1}$, the utility for an arm $i \in I_{-1}$ is simulated by returning $1-x_t$ (flipping the value of the next available sample).

    \cref{algorithm:coin-reduction} alternates between simulating the interaction of $(\verifier_{1},\prover_{1})$ and the interaction of $(\verifier_{-1},\prover_{-1})$. Each interaction is simulated until a sample $x_t$ is used, at which point \cref{algorithm:coin-reduction} switches to simulating the other interaction. This ensures that the numbers of samples $x_t$ used for the two simulations differ by at most $1$. 

    Assume that the bias of the coin is $1/2 +b^*\cdot\varepsilon$ for some fixed but unknown $b^* \in \set{-1,1}$. 
    Observe that for all $b \in \set{-1,1}$, all arms in the bandit oracle for $\verifier_b$ that are not in $I_b$ have utility distribution $\Ber{1/2-\varepsilon}$; however, in simulation $(\verifier_{b^*},\prover_{b^*})$, the arms in $I_{b^*}$ have utility distribution $\Ber{1/2+(b^*)(b^*)\cdot\varepsilon} = \Ber{1/2+\varepsilon}$, while for $(\verifier_{-b^*},\prover_{-b^*})$ the arms in $I_{-b^*}$ have utility distribution $\Ber{1/2+(-b^*)(b^*)\cdot\varepsilon} = \Ber{1/2-\varepsilon}$. In particular, for exactly one of the simulations, the expected utility of \emph{all} $n$ arms the verifier's oracle is $1/2-\varepsilon$, and in the other simulation there is a randomly-chosen subset of $1/\sigma$ arms that have expected utility $1/2 +\varepsilon$, and the remaining $n-1/\sigma$ arms have expected utility $1/2 - \varepsilon$.

    We now show that \cref{algorithm:coin-reduction} solves the coin bias problem correctly. %
    In \cref{algorithm:coin-reduction}, either the simulation of $(\verifier_{b^*},\prover_{b^*})$ or of $(\verifier_{-b^*},\prover_{-b^*})$ terminates first and determines the output of \cref{algorithm:coin-reduction}. If $(\verifier_{b^*},\prover_{b^*})$ terminates first, then the soundness of $(\verifier_{b^*},\prover_{b^*})$ implies that with probability at least~$2/3$, $\verifier_{b^*}$ outputs $\pi$ such that $\pi = \reject$ or $\pi$ is a $\sigma$-smooth $\varepsilon$-optimal policy for the bandit that $\verifier_{b^*}$ had oracle access to, which is an oracle where each arm $i \in [n]$ has expected utility 
    \[
        u_i = 1/2-(-1)^{\indicator{i \in I_{b^*}\!}}\cdot\varepsilon.
    \]
    The optimal $\sigma$-smooth policy for that bandit is given by $\pi^*_i = \indicator{i \in I_{b^*}}\cdot\sigma$, which has expected utility $\pi^* \cdot u = 1/2+\varepsilon$. Thus, with probability at least~$2/3$,
    \begin{equation}
        \label{eq:pi-utility-at-least-half}
        \pi \cdot u \geq \pi^* \cdot u - \varepsilon = \frac{1}{2}.
    \end{equation}
    But 
    \begin{equation}
        \label{eq:pi-utility-breakdown}
        \pi \cdot u = \paren*{\frac{1}{2}+\varepsilon}\cdot\sum_{i \in I_{b^*}} \pi_i + \paren*{\frac{1}{2}-\varepsilon}\cdot\sum_{i \notin I_{b^*}} \pi_i.
    \end{equation}
    Combining \cref{eq:pi-utility-at-least-half,eq:pi-utility-breakdown} gives that with probability at least~$2/3$,
    \begin{eqnarray}
        \sum_{i \in I_{b^*}}\pi_i \geq \frac{1}{2}.
    \end{eqnarray}
    Therefore, if $(\verifier_{b^*},\prover_{b^*})$ terminates first, then with probability at least~$2/3$, $\verifier_{b^*}$ reaches the \codehyperlink{line:return-b}{first output statement} in \cref{algorithm:coin-reduction} and outputs $b^*$, representing a coin bias of $\Ber{1/2+b^*\cdot\varepsilon}$, which is the correct answer.
    
    On the other hand, in the copy $(\verifier_{-b^*},\prover_{-b^*})$, all the arms in the verifier's bandit oracle and in the prover's bandit oracle have expected utility $1/2 - \varepsilon$. By completeness of $(\verifier_{-b^*},\prover_{-b^*})$, with probability at least~$2/3$, the verifier $\verifier_{-b^*}$ outputs a policy $\pi\neq\reject$ that is a distribution over~$[n]$. Because in that simulation all the arms in~$[n]$ are indistinguishable, and the set $I_{-b^*}$ is chosen uniformly at random, the expected weight that $\pi$ assigns to arms in $I_{-b^*}$ is
    \[
        \EE{\sum_{i \in I_{-b^*}} \pi_{i}} = \frac{\abs*{I_{-b^*}}}{n} = \frac{\paren*{\frac{1}{\sigma}}}{n} = \frac{1}{\sigma n}.
    \]
    By Markov's inequality,
    \begin{equation}
        \PP{
            \sum_{i \in I_{-b^*}} \pi_{i} \geq \frac{1}{2}
        }
        \leq \frac{2}{n\sigma} \leq \frac{1}{12},
    \end{equation}
    where we have used the assumption that $\sigma \geq 24/n$.
    Therefore, if $(\verifier_{-b^*},\prover_{-b^*})$ terminates first, then with probability at least 
    \[
    \frac{2}{3}-\frac{1}{12} = \frac{7}{12},
    \] 
    $\verifier_{-b^*}$ reaches the \codehyperlink{line:return-minus-b}{second output statement} in \cref{algorithm:coin-reduction} and outputs $-(-b^*) = b^*$, which again is the correct answer. This establishes that \cref{algorithm:coin-reduction} solves the coin bias problem correctly with probability at least~$7/12$.

    We now show that the correctness of \cref{algorithm:coin-reduction} implies a lower bound on the number of oracle queries used by the verifier in bandit verification. Let $m_{b^*}$ and $m_{-b^*}$ be the number of coin samples $x_t$ used by $\verifier_{b^*}$ and $\verifier_{-b^*}$, respectively. Let $k$ be the total number of oracle queries performed by $\verifier_{-b^*}$ (some of which used coin samples, so $m_{-b^*} \leq k$). We want to show a lower bound on $k$.

    Seeing as \cref{algorithm:coin-reduction} solves the coin bias problem correctly with probability at least $7/12$, \cref{claim:lower-bound-for-coins} implies that there exists a constant $c_0>0$ such that the total number $m_{b^*} + m_{-b^*}$ of coin samples used by \cref{algorithm:coin-reduction} is lower bounded by
    \begin{equation}
        \label{eq:coins-used-is-large}
        m_{b^*} + m_{-b^*} \geq c_0 \cdot \frac{1}{\varepsilon^2}.
    \end{equation} 
    In the simulation $(\verifier_{-b^*},\prover_{-b^*})$, all the arms in the bandit oracles for $\verifier_{-b^*}$ and $\prover_{-b^*}$ have expected utility $1/2 - \varepsilon$. For these oracles, the arms in $I_{-b^*}$ are indistinguishable from the other arms in~$[n]$. Hence, if $\verifier_{-b^*}$ makes $k$ queries to the bandit oracle, then the expected number queries that $\verifier_{-b^*}$ makes to arms in $I_{-b^*}$, and therefore the number of a coin samples $\verifier_{-b^*}$ uses, is
    \[
        \EE{m_{-b^*}} = k \cdot \frac{|I_{-b^*}|}{n} = k \cdot \frac{\paren*{\frac{1}{\sigma}}}{n} = \frac{k}{\sigma n}.
    \]
    By Markov's inequality,
    \[
        \PP{m_{-b^*} \geq 10\cdot\frac{k}{ \sigma n}} \leq \frac{1}{10}.
    \]
    In other words, with probability at least $9/10$,
    \begin{equation}
        \label{eq:coins-used-per-simualtion}
        m_{-b^*} < 10\cdot\frac{k}{ \sigma n}.
    \end{equation}
    Because \cref{algorithm:coin-reduction} alternates between simulating $(\verifier_{b^*},\prover_{b^*})$ and $(\verifier_{-b^*},\prover_{-b^*})$, the numbers $m_{b^*}$ and $m_{-b^*}$ of coin samples used by each simulation differ by at most $1$. Hence, 
    \begin{equation}
        \label{eq:coins-used-is-small}
        m_{b^*} + m_{-b^*} \leq 2m_{-b^*}+1.
    \end{equation}
    We conclude that 
    \begin{align*}
        k
        &>
        \frac{n\sigma}{10} \cdot m_{-b^*}
        \tagexplain{By \cref{eq:coins-used-per-simualtion}}
        \\
        &\geq
        \frac{n\sigma}{10} \cdot \frac{%
            m_{b^*}
            +
            m_{-b^*}-1
        }{2}
        \tagexplain{By \cref{eq:coins-used-is-small}}
        \\
        &\geq
        \frac{n\sigma}{10} \cdot \frac{%
            \frac{c_0}{\varepsilon^2}-1
        }{2}
        \tagexplain{By \cref{eq:coins-used-is-large}}
        \\
        &=
        \frac{c_0}{20}\cdot\frac{n\sigma}{\varepsilon^2} - \frac{1}{20},
    \end{align*}
    as desired.
\end{proofof}

%% file: parts/appendix/bandits-proofs/lower-bound-for-learning.tex
\subsection{Proof of lower bound for learning smooth bandit strategies}
\label{appendix:bandit-learning-lower-bound}

\begin{proofof}{\cref{claim:lower-bound-learning-smooth-bandit}}
    For simplicity, we assume that $1/\sigma$ is an integer (the proof for the general case is similar). We will assume in the proof that $m \leq n/2$ (otherwise, there is nothing to prove).
    
    For each $b\in \{0,1\}$, let $\cD_b$ denote the degenerate distribution such that $\PPP{x \sim \cD_b}{x = b} = 1$.
    Let $\cS_\sigma = \binom{[n]}{1/\sigma}$ be the collection of all subsets of $[n]$ of size $1/\sigma$. For every set $S \in \cS_\sigma$, let $q^S \in \left(\distribution{[0,1]}\right)^n$ be an $n$-armed bandit such that for each $i \in [n]$, $q^S_i = \cD_{\indicator{i \in S}}$. In words, $q^S$ is a bandit where arms in $S$ always give utility $1$, and the remaining arms always give utility $0$. Let $u^S = \utility{q^S}$ denote the expected utilities vector of the bandit $q^S$. For $S \in \cS_\sigma$ and $i \in [n]$, let $S_i = \indicator{i \in S}$.

    For any $S \in \cS_\sigma$, let $\pi^S = (\pi^S_1,\dots,\pi^S_n)$ be the uniform distribution on $S$, i.e., $\pi^S_i = \1(i \in S)\sigma$. Note that $\pi^S$ is a $\sigma$-smooth strategy, and it has expected utility 
    \begin{equation}
        \label{eq:optimal-smoooth-utility}
        \pi^S \cdot u^S = \sum_{i \in S} \sigma \cdot 1 = 1.
    \end{equation}
    Consider the following experiment:
    \begin{enumerate}
        \item{
            Sample a subset $S$ uniformly at random from $\cS_\sigma$.
        }
        \item{
            Execute $A$ with respect to the bandit $q^S$.
        }
    \end{enumerate}
    
    Let $I = \{I_1,\dots,I_m\}$ be the indices of the arms pulled by $A$, and let $G = S \cap I$ be the ``good'' arms pulled by $A$ (i.e., the arms queried by $A$ that have utility $1$).
    
    In the experiment, for each $i \in [n]$, $\EE{S_i} = |S|/n = 1/(\sigma n)$. We may assume without loss of generality that $A$ issues precisely $m$ queries, each to a different arm.\footnote{%
        In the experiment, the distributions of each arm is degenerate. So an algorithm that issues less than $m$ queries, or queries the same arm more than once, can be transformed into an algorithm with the same output that pulls precisely $m$ distinct arms.%
    } 
    It follows that
    \begin{equation}
        \label{eq:expected-size-of-G}
        \EE{|G|}
        =
        \EE{\sum_{t = 1}^m S_{I_t}}
        =
        \sum_{t=1}^m \EE{S_{I_t}}
        =
        m\cdot\EE{S_1}
        =
        \frac{m}{\sigma n}.
    \end{equation}
    
    The strategy $\pi$ that $A$ outputs in the experiment has expected utility
    \begin{equation}
        \label{eq:expected-utility-A}
        \EE{
            \pi \cdot u^S
        }
        =
        \EE{
            \sum_{i = 1}^n \pi_i \cdot u^S_i
        }
        =
        \EE{
            \sum_{i \in G} \pi_i \cdot u^S_i
        } + 
        \EE{
            \sum_{i \in [n] \setminus G} \pi_i \cdot u^S_i
        }.
    \end{equation}
    Consider each term separately. For the first term,
    \begin{equation}
        \label{eq:expected-utility-G}
        \EE{
            \sum_{i \in G} \pi_i \cdot u^S_i
        }
        \leq
        \EE{
            \sum_{i \in G} \pi_i \cdot 1
        }
        =
        \EE{
            \pi_G
        },
    \end{equation}
    where $\pi_G = \sum_{i \in G} \pi_i$. 
    For the second term, 
    \begin{align}
        \label{eq:expected-utility-not-G}
        \EE{
            \sum_{i \in [n] \setminus G} \pi_i \cdot  u^S_i
        }
        &=
        \EE{
            \sum_{i \in [n] \setminus G} \pi_i \cdot S_i
        }
        \nonumber
        \\
        &\leq
        \frac{1}{\sigma (n-m)} \cdot \EE{
            \sum_{i \in [n] \setminus G} \pi_i
        }
        \nonumber
        \\
        &\leq
        \frac{2}{\sigma n} \cdot \EE{
            1 - \pi_G
        }.
        \tagexplain{$m \leq n/2$}
    \end{align}

    Combining \cref{eq:expected-utility-A,eq:expected-utility-G,eq:expected-utility-not-G} yields
    \[
        \EE{
            \pi \cdot u^S
        }
        \leq
        \EE{
             \pi_G + \frac{2}{\sigma n} \cdot (1-\pi_G)
        }
        =
        \frac{2}{\sigma n} + \left(1-\frac{2}{\sigma n}\right)\EE{
            \pi_G
        }.
    \]
    From \cref{eq:expected-size-of-G} and the $\sigma$-smoothness of $\pi$,  
    \[
        \EE{\pi_G} = \EE{\sum_{i \in G} \pi_i} \leq \EE{\sigma \cdot |G|} = \frac{m}{n}.
    \]
    Namely, 
    \begin{equation}
        \label{eq:upper-bound-A-utility}
        \EE{
            \pi \cdot u^S
        }
        \leq 
        \frac{2}{\sigma n} + \left(1-\frac{2}{\sigma n}\right)
        \cdot
        \frac{m}{n}.
    \end{equation}
    
    From the utility of the optimal $\sigma$-smooth strategy (\cref{eq:optimal-smoooth-utility}) and the assumption that with probability at least $2/3$, $A$ outputs an $\varepsilon$-optimal $\sigma$-smooth strategy,
    \[
        \PP{\pi \cdot u^S \geq 1-\varepsilon} \geq 2/3,
    \]
    so
    \begin{equation}
        \label{eq:lower-bound-A-utility}
        \EE{
            \pi \cdot u^S
        }
        \geq 
        \PP{\pi \cdot u^S \geq 1-\varepsilon}\cdot(1-\varepsilon)
        \geq
        \frac{2}{3}\cdot(1-\varepsilon)
        \geq
        \frac{1}{2}.
    \end{equation}

    Combining \cref{eq:upper-bound-A-utility,eq:lower-bound-A-utility} yields,
    \[
        \frac{2}{\sigma n} + \left(1-\frac{2}{\sigma n}\right)
        \cdot
        \frac{m}{n}
        \geq
        \frac{1}{2}.
    \]
    We conclude that 
    \begin{align*}
        m 
        &\geq
        n \cdot \frac{\sigma n}{\sigma n-2} \cdot \left(
            \frac{1}{2} - \frac{2}{\sigma n}
        \right)
        \geq
        \frac{n}{6}.
        \tagexplain{$\sigma \geq 5/n$}
    \end{align*}
\end{proofof}

%% file: parts/appendix/games-proofs/games-proofs.tex
\section{Proofs for games}

\input{parts/appendix/games-proofs/game-verification-protocol}

\input{parts/appendix/games-proofs/game-verification-lower-bound}

%% file: parts/appendix/games-proofs/game-verification-protocol.tex
\subsection{Proof for game verification} \label{appendix:game-verification-protocol}

In this appendix we prove \cref{theorem:game-verification}.

\begin{proofof}{\cref{theorem:game-verification}}
    \cref{protocol:game-verification} reduces the problem of verifying approximate optimality of a smooth Nash equilibrium to several bandit verification tasks, one for each player.
    
    That is, verifying that each player $i \in [k]$ has no deviation to a $\sigma$-smooth strategy increasing its expected utility by at least $\varepsilon$ is equivalent to verifying a bandit problem defined as follows.
    Observe that for each action $j \in [n]$, $\pi$ specifies a distribution over player $i$'s utility, given by the output of the game oracle $\cO_u(\pi)_i$.
    This induces an $n$-arm bandit, denoted $\cB(i, u, \pi)$.
    A strategy $\pi_i$ for this bandit is $\sigma$-smooth if and only if $\pi_i$ is a $\sigma$-smooth strategy for player $i$ in the given game.
    Therefore, player $i$ has no smooth deviation increasing their utility by at least $\varepsilon$ under $\pi$ if and only if $\pi_i$ is an $\varepsilon$-approximate $\sigma$-smooth policy for $\cB(i, u, \pi)$.
    
	\cref{protocol:game-verification} leverages this equivalence between each player's optimality in the game and bandit optimality, simply by having the prover and verifier engage in a verification protocol for each of these bandits.
	The prover sends $k$ length-$n$ vectors of empirical average rewards $\tilde{t}_i$, one for each bandit.
	The verifier then checks optimality of $\pi$ for player $i$ under bandit $\cB(i, u, \pi)$ with maximum error probability $\delta = 1/(3k)$.
	If any of these subprotocols fails, the verifier rejects and terminates.
	
	\paragraph{Completeness.} Observe that if $\pi$ is indeed an $\varepsilon$-approximately optimal $\sigma$-smooth strategy for the game, every induced bandit policy is approximately optimal as well.
	By \Cref{lemma:verify-smooth-bandit}, for each $i$ \Cref{algorithm:optimal-smooth-bandit-strategy} succeeds with probability at least $1/3k$. 
	By a union bound, all $k$ subprotocols succeed with probability at least $2/3$.

	\paragraph{Soundness.} If $\pi$ is not an $(\varepsilon + \eta)$-approximately optimal smooth strategy, there must be a player $i$ for which the induced bandit policy is not approximately optimal.
	For this player, the bandit verification subprotocol rejects with probability at least $1 - 1/3k \geq 2/3$.

	\paragraph{Query complexity.}
	The prover and verifier engage in $k$ runs of \Cref{protocol:verify-bandit-variant} with $\delta = 1/3k$.
	The query complexity follows.
\end{proofof}

%% file: parts/appendix/games-proofs/game-verification-lower-bound.tex
\subsection{Proof of lower bound for smooth Nash verification} \label{appendix:game-verification-lower-bound}

\begin{proofof}{\cref{theorem:game-verification-lower-bound}}
    For simplicity, assume that $1/\sigma$ is an integer.  
    Let $\cS = \binom{[n]}{1/\sigma}$ be the collection of all subsets of $[n]$ of size $1/\sigma$. For every vector $s \in \cS^k$ and integer $i^* \in [k]$, let $u^{s,i^*} = (u^{s,i^*}_1,\dots,u^{s,i^*}_k)$ be a $k$-player $n$-action game as follows. For each $i \in k$, $u^{s,i^*}_i: ~ [n]^k \to [0,1]$ is a utility functions such that for every action vector $a \in [n]^k$,
    \[
        u^{s,i^*}_i(a) =
        \left\{
        \begin{array}{ll}
        1  & i = i^* ~ \land ~ \left(\forall j \in [k]: ~ a_j \in s_j\right)  \\
        0 & \textrm{otherwise}.
        \end{array}
        \right.
    \]
    Note that there exists a $\sigma$-smooth profile strategy $\pi^{s,i^*}$ for $u^{s,i^*}$ where player $i^*$ has expected utility $1$, and all other players have expected utility $0$. Namely, $\pi^{s,i^*}$ is the profile where each player $i \in [k]$ selects an action uniformly at random from the set $s_i \in \cS$. 

    For every vector $s \in \cS^k$ and integer $i^* \in [k]$, define a distribution $\cD^{s,i^*}$ over strategy profiles as follows. When $\pi \sim \cD^{s,i^*}$, then with probability $1$, for every $i \in [k] \setminus \{i^*\}$, $\pi_i = \uniform{s_i}$ (as in the strategy $\pi^{s,i^*}$ described in the previous paragraph).

    For player $i^*$, the strategy $\pi_{i^*}$ is sampled as follows.
    \begin{enumerate}
        \item{
            Sample $R \sim \uniform{\cS}\: \vert \: \left(R \cap s_{i^*} = \varnothing\right)$.
        }
        \item{
            Set $\pi_{i^*} = \uniform{R}$.
        }
    \end{enumerate}

    Let $u^0 = (u^0_1,\dots,u^0_k)$ denote the game where all players always receive utility $0$ ($u^0_i \equiv 0$ is a constant function for all $i \in [k]$).

    Now, consider the following experiment.
    \begin{enumerate}
        \item{
            Sample $i^* \sim \uniform{[k]}$
        }
        \item{
            Sample $s \sim \uniform{\cS^k}$
        }
        \item{
            Sample $\pi \sim \cD^{s,i^*}$.
        }
        \item{
            Sample a bit $b \sim \uniform{\{0,1\}}$
        }
        \item{
            Execute the honest prover $P(k,n,\varepsilon,\sigma,\eta,\pi)$ with access to the game oracle $\cO_{u^0}$.
        }
        \item{
            If $b = 0$, execute the verifier $\verifier(k,n,\varepsilon,\sigma,\eta,\pi)$ with access to the game oracle $\cO_{u^0}$. Otherwise, if $b = 1$, execute the verifier $\verifier(k,n,\varepsilon,\sigma,\eta,\pi)$ with access to the game oracle $\cO_{u^{s,i^*}}$.
        }
    \end{enumerate}

    Observe that if $b=0$, then $\verifier$ should accept, because in the game $u^0$, every action profile is a Nash equilibrium (and in addition, the profile $\pi$ sampled from $\cD^{s,i^*}$ is also $\sigma$-smooth). On the other hand, when $b = 1$ then $\verifier$ should usually reject, because $\pi$ will have expected utility close to $0$ for all players, but player $i^*$ has a deviation that would give it an expected utility of $1$.

    We now argue that $\verifier$ cannot distinguish between the case $b=0$ and $b=1$ unless it makes at least $\OmegaOf{kn\sigma}$ queries to the game oracle. 
    
    To see this, make a few observations:
    \begin{itemize}
        \item{
            Without loss of generality, we can assume that every query $\pi'$ that $\verifier$ sends to the game oracle is a strategy profile where each player $i \in [k]$ playes a pure strategy. (If $\verifier$ wants to sends a query that includes a mixed strategy, then $\verifier$ can simluate the result of that query by first sampling an action vector $a \sim \pi'$, and then sending the query consisting of a pure profile corresponding to $a$ to the oracle. The result is identical.)
        }
        \item{
            Every action vector $a \in [n]^k$ where the number of deviations is not $1$, namely,
            \[
            \left|
                \big\{
                    i \in [k]: ~ a_i \notin \supp (\pi_i)
                \big\}
            \right| \neq 1,
            \]
            satisfies that $u^{s,i^*}(a) = \mathbf{0} = u^{0}(a)$, where $\mathbf{0} = (0,\dots,0)$. Hence, we may assume without loss of generality that each query that $\verifier$ sends to the game oracle has precisely one player $i \in [k]$ that deviates to an action not in $\supp(\pi_i)$.
        }
        \item{
            Assume $b = 1$. Fix an action vector $a \in [n]^k$.\footnote{%
                $a$ is simply a fixed action vector. In particular, it does not depend on any of the random variables in the experiment. 
            } If there exists a unique $j \in [k]$ such that $a_j \notin \supp(\pi_j)$, then the probability that the result of the query $\pi_a$ corresponding to action profile $a$ is not $\mathbf{0}$ is
            \begin{align}
                \label{eq:query-success-rate}
                \alpha :&= \PP{\cO_{u^{s,i^*}}(\pi_a) \neq \mathbf{0}}
                \\
                &= 
                \PP{
                    j = i^* ~ \land ~ a_j \in s_j
                }
                \nonumber
                \\
                &= 
                \PP{
                    j = i^*
                }
                \cdot
                \PP{
                    a_j \in s_j ~ | ~ j = i^*
                }
                \nonumber
                \\
                &=
                \frac{1}{k}\cdot\frac{\frac{1}{\sigma}}{n-\frac{1}{\sigma}} \leq \frac{2}{kn\sigma}.
                \tagexplain{$\sigma \geq 2/n$}
                \nonumber
            \end{align}
        }
    \end{itemize}
    From \cref{eq:query-success-rate}, it follows that the number $N$ of queries that $\verifier$ sends before it has its first success (i.e., the number of queries that receives result $\mathbf{0}$ prior to the first query that receives a nonzero result), is distributed geometrically, ${N \sim \Geom{\alpha}}$. In particular,
    \begin{align*}
        \PP{N \geq m} = \left(1-\alpha\right)^m.
    \end{align*}
    Assume for contradiction that $\verifier$ performs at most $\mV$ queries and has success probability $2/3$. Then for some $c \in [0,1)$,
    \[
        1 
        > 
        c 
        \geq 
        \PP{N \geq \mV}
        = \left(1-\alpha\right)^{\mV}
        \geq 
        \left(e^{-\frac{\alpha}{1-\alpha}}\right)^{\mV}.
    \]
    This implies that 
    \[
        \mV \geq \OmegaOf{\frac{1-\alpha}{\alpha}} = \OmegaOf{\frac{1}{\alpha}} = \OmegaOf{kn\sigma},
    \]
    as desired.
\end{proofof}

%% file: parts/appendix/misc-appendices.tex
\section{Concentration of measure}

\begin{theorem}[\citealp{hoeffding1963probability}]
	\label{theorem:hoeffding}
	Let $a,b,\mu \in \bbR$ and $m \in \bbN$. Let $Z_1, \dots, Z_m$ be a sequence of i.i.d.\ real-valued random variables and let $Z=\frac{1}{m} \sum_{i=1}^m Z_i$. Assume that $\EE{Z}=\mu$, and for every $i \in [m]$, $\PP{a \leq Z_i \leq b}=1$. Then, for every $\varepsilon>0$,
	\[
		\PP{\left|Z-\mu\right|>\varepsilon} \leq 2 \expf{\frac{-2 m \varepsilon^2}{(b-a)^2}}.
	\]
\end{theorem}

%% file: arXiv/main.bbl
\begin{thebibliography}{35}
\providecommand{\natexlab}[1]{#1}
\providecommand{\url}[1]{\texttt{#1}}
\expandafter\ifx\csname urlstyle\endcsname\relax
  \providecommand{\doi}[1]{doi: #1}\else
  \providecommand{\doi}{doi: \begingroup \urlstyle{rm}\Url}\fi

\bibitem[Anthony and Bartlett(2002)]{DBLPbooksdaglib0025992}
Martin Anthony and Peter~L. Bartlett.
\newblock \emph{Neural Network Learning - Theoretical Foundations}.
\newblock Cambridge University Press, 2002.
\newblock ISBN 978-0-521-57353-5.
\newblock URL \url{http://www.cambridge.org/gb/knowledge/isbn/item1154061/?site\_locale=en\_GB}.

\bibitem[Assadi and Wang(2022)]{DBLPconfnipsAssadi022}
Sepehr Assadi and Chen Wang.
\newblock Single-pass streaming lower bounds for multi-armed bandits exploration with instance-sensitive sample complexity.
\newblock In Sanmi Koyejo, S.~Mohamed, A.~Agarwal, Danielle Belgrave, K.~Cho, and A.~Oh, editors, \emph{Advances in Neural Information Processing Systems 35: Annual Conference on Neural Information Processing Systems 2022, NeurIPS 2022, New Orleans, LA, USA, November 28 - December 9, 2022}, 2022.
\newblock URL \url{http://papers.nips.cc/paper\_files/paper/2022/hash/d5e9cf50dc182447a916bc56d4d42942-Abstract-Conference.html}.

\bibitem[Babichenko(2019)]{DBLPjournalssigecomBabichenko19}
Yakov Babichenko.
\newblock Informational bounds on equilibria (a survey).
\newblock \emph{SIGecom Exch.}, 17\penalty0 (2):\penalty0 25--45, 2019.
\newblock \doi{10.1145/3381329.3381333}.
\newblock URL \url{https://doi.org/10.1145/3381329.3381333}.

\bibitem[Canetti and Karchmer(2021)]{DBLPconftccCanettiK21}
Ran Canetti and Ari Karchmer.
\newblock Covert learning: How to learn with an untrusted intermediary.
\newblock In Kobbi Nissim and Brent Waters, editors, \emph{Theory of Cryptography - 19th International Conference, {TCC} 2021, Raleigh, NC, USA, November 8-11, 2021, Proceedings, Part {III}}, volume 13044 of \emph{Lecture Notes in Computer Science}, pages 1--31. Springer, 2021.
\newblock \doi{10.1007/978-3-030-90456-2\_1}.
\newblock URL \url{https://doi.org/10.1007/978-3-030-90456-2\_1}.

\bibitem[Caro et~al.(2024{\natexlab{a}})Caro, Eisert, Hinsche, Ioannou, Nietner, and Sweke]{DBLPjournalscorrabs241023969}
Matthias~C. Caro, Jens Eisert, Marcel Hinsche, Marios Ioannou, Alexander Nietner, and Ryan Sweke.
\newblock Interactive proofs for verifying (quantum) learning and testing.
\newblock \emph{CoRR}, abs/2410.23969, 2024{\natexlab{a}}.
\newblock \doi{10.48550/ARXIV.2410.23969}.
\newblock URL \url{https://doi.org/10.48550/arXiv.2410.23969}.

\bibitem[Caro et~al.(2024{\natexlab{b}})Caro, Hinsche, Ioannou, Nietner, and Sweke]{DBLPconfinnovationsCaroHINS24}
Matthias~C. Caro, Marcel Hinsche, Marios Ioannou, Alexander Nietner, and Ryan Sweke.
\newblock Classical verification of quantum learning.
\newblock In Venkatesan Guruswami, editor, \emph{15th Innovations in Theoretical Computer Science Conference, {ITCS} 2024, January 30 to February 2, 2024, Berkeley, CA, {USA}}, volume 287 of \emph{LIPIcs}, pages 24:1--24:23. Schloss Dagstuhl - Leibniz-Zentrum f{\"{u}}r Informatik, 2024{\natexlab{b}}.
\newblock \doi{10.4230/LIPICS.ITCS.2024.24}.
\newblock URL \url{https://doi.org/10.4230/LIPIcs.ITCS.2024.24}.

\bibitem[Catalano and Fiore(2013)]{DBLPconfpkcCatalanoF13}
Dario Catalano and Dario Fiore.
\newblock Vector commitments and their applications.
\newblock In Kaoru Kurosawa and Goichiro Hanaoka, editors, \emph{Public-Key Cryptography - {PKC} 2013 - 16th International Conference on Practice and Theory in Public-Key Cryptography, Nara, Japan, February 26 - March 1, 2013. Proceedings}, volume 7778 of \emph{Lecture Notes in Computer Science}, pages 55--72. Springer, 2013.
\newblock \doi{10.1007/978-3-642-36362-7\_5}.
\newblock URL \url{https://doi.org/10.1007/978-3-642-36362-7\_5}.

\bibitem[Charikar et~al.(2017)Charikar, Steinhardt, and Valiant]{DBLPconfstocCharikarSV17}
Moses Charikar, Jacob Steinhardt, and Gregory Valiant.
\newblock Learning from untrusted data.
\newblock In Hamed Hatami, Pierre McKenzie, and Valerie King, editors, \emph{Proceedings of the 49th Annual {ACM} {SIGACT} Symposium on Theory of Computing, {STOC} 2017, Montreal, QC, Canada, June 19-23, 2017}, pages 47--60. {ACM}, 2017.
\newblock \doi{10.1145/3055399.3055491}.
\newblock URL \url{https://doi.org/10.1145/3055399.3055491}.

\bibitem[Chen and Li(2015)]{DBLPjournalscorrChenL15c}
Lijie Chen and Jian Li.
\newblock On the optimal sample complexity for best arm identification.
\newblock \emph{CoRR}, abs/1511.03774, 2015.
\newblock URL \url{http://arxiv.org/abs/1511.03774}.

\bibitem[Chen et~al.(2017)Chen, Li, and Qiao]{DBLPconfcoltChenLQ17}
Lijie Chen, Jian Li, and Mingda Qiao.
\newblock Towards instance optimal bounds for best arm identification.
\newblock In Satyen Kale and Ohad Shamir, editors, \emph{Proceedings of the 30th Conference on Learning Theory, {COLT} 2017, Amsterdam, The Netherlands, 7-10 July 2017}, volume~65 of \emph{Proceedings of Machine Learning Research}, pages 535--592. {PMLR}, 2017.
\newblock URL \url{http://proceedings.mlr.press/v65/chen17b.html}.

\bibitem[Chen and Deng(2006)]{DBLPconffocsChenD06}
Xi~Chen and Xiaotie Deng.
\newblock Settling the complexity of two-player nash equilibrium.
\newblock In \emph{47th Annual {IEEE} Symposium on Foundations of Computer Science {(FOCS} 2006), 21-24 October 2006, Berkeley, California, USA, Proceedings}, pages 261--272. {IEEE} Computer Society, 2006.
\newblock \doi{10.1109/FOCS.2006.69}.
\newblock URL \url{https://doi.org/10.1109/FOCS.2006.69}.

\bibitem[Cover and Thomas(2006)]{DBLPbooksdaglib0016881}
Thomas~M. Cover and Joy~A. Thomas.
\newblock \emph{Elements of Information Theory, 2nd Edition}.
\newblock Wiley, 2006.
\newblock ISBN 978-0-471-24195-9.
\newblock URL \url{http://www.elementsofinformationtheory.com/}.

\bibitem[Daskalakis et~al.(2009)Daskalakis, Goldberg, and Papadimitriou]{DBLPjournalscacmDaskalakisGP09}
Constantinos Daskalakis, Paul~W. Goldberg, and Christos~H. Papadimitriou.
\newblock The complexity of computing a nash equilibrium.
\newblock \emph{Commun. {ACM}}, 52\penalty0 (2):\penalty0 89--97, 2009.
\newblock \doi{10.1145/1461928.1461951}.
\newblock URL \url{https://doi.org/10.1145/1461928.1461951}.

\bibitem[Daskalakis et~al.(2024)Daskalakis, Golowich, Haghtalab, and Shetty]{DBLPconfinnovationsDaskalakisGHS24}
Constantinos Daskalakis, Noah Golowich, Nika Haghtalab, and Abhishek Shetty.
\newblock Smooth {N}ash equilibria: Algorithms and complexity.
\newblock In Venkatesan Guruswami, editor, \emph{15th Innovations in Theoretical Computer Science Conference, {ITCS} 2024, January 30 to February 2, 2024, Berkeley, CA, {USA}}, volume 287 of \emph{LIPIcs}, pages 37:1--37:22. Schloss Dagstuhl - Leibniz-Zentrum f{\"{u}}r Informatik, 2024.
\newblock \doi{10.4230/LIPICS.ITCS.2024.37}.
\newblock URL \url{https://doi.org/10.4230/LIPIcs.ITCS.2024.37}.

\bibitem[Even{-}Dar et~al.(2002)Even{-}Dar, Mannor, and Mansour]{DBLPconfcoltEvenDarMM02}
Eyal Even{-}Dar, Shie Mannor, and Yishay Mansour.
\newblock {PAC} bounds for multi-armed bandit and markov decision processes.
\newblock In Jyrki Kivinen and Robert~H. Sloan, editors, \emph{Computational Learning Theory, 15th Annual Conference on Computational Learning Theory, {COLT} 2002, Sydney, Australia, July 8-10, 2002, Proceedings}, volume 2375 of \emph{Lecture Notes in Computer Science}, pages 255--270. Springer, 2002.
\newblock \doi{10.1007/3-540-45435-7\_18}.
\newblock URL \url{https://doi.org/10.1007/3-540-45435-7\_18}.

\bibitem[Fearnley et~al.(2015)Fearnley, Gairing, Goldberg, and Savani]{DBLPjournalsjmlrFearnleyGGS15}
John Fearnley, Martin Gairing, Paul~W. Goldberg, and Rahul Savani.
\newblock Learning equilibria of games via payoff queries.
\newblock \emph{J. Mach. Learn. Res.}, 16:\penalty0 1305--1344, 2015.
\newblock \doi{10.5555/2789272.2886792}.
\newblock URL \url{https://dl.acm.org/doi/10.5555/2789272.2886792}.

\bibitem[Goldwasser et~al.(1989)Goldwasser, Micali, and Rackoff]{DBLPjournalssiamcompGoldwasserMR89}
Shafi Goldwasser, Silvio Micali, and Charles Rackoff.
\newblock The knowledge complexity of interactive proof systems.
\newblock \emph{{SIAM} J. Comput.}, 18\penalty0 (1):\penalty0 186--208, 1989.
\newblock \doi{10.1137/0218012}.
\newblock URL \url{https://doi.org/10.1137/0218012}.

\bibitem[Goldwasser et~al.(2015)Goldwasser, Kalai, and Rothblum]{DBLPjournalsjacmGoldwasserKR15}
Shafi Goldwasser, Yael~Tauman Kalai, and Guy~N. Rothblum.
\newblock Delegating computation: Interactive proofs for muggles.
\newblock \emph{J. {ACM}}, 62\penalty0 (4):\penalty0 27:1--27:64, 2015.
\newblock \doi{10.1145/2699436}.
\newblock URL \url{https://doi.org/10.1145/2699436}.

\bibitem[Goldwasser et~al.(2021)Goldwasser, Rothblum, Shafer, and Yehudayoff]{DBLPconfinnovationsGoldwasserRSY21}
Shafi Goldwasser, Guy~N. Rothblum, Jonathan Shafer, and Amir Yehudayoff.
\newblock Interactive proofs for verifying machine learning.
\newblock In James~R. Lee, editor, \emph{12th Innovations in Theoretical Computer Science Conference, {ITCS} 2021, January 6-8, 2021, Virtual Conference}, volume 185 of \emph{LIPIcs}, pages 41:1--41:19. Schloss Dagstuhl - Leibniz-Zentrum f{\"{u}}r Informatik, 2021.
\newblock \doi{10.4230/LIPICS.ITCS.2021.41}.
\newblock URL \url{https://doi.org/10.4230/LIPIcs.ITCS.2021.41}.

\bibitem[G{\"{o}}{\"{o}}s and Rubinstein(2023)]{DBLPjournalssiamcompGoosR23}
Mika G{\"{o}}{\"{o}}s and Aviad Rubinstein.
\newblock Near-optimal communication lower bounds for approximate nash equilibria.
\newblock \emph{{SIAM} J. Comput.}, 52\penalty0 (6):\penalty0 S18--316, 2023.
\newblock \doi{10.1137/19M1242069}.
\newblock URL \url{https://doi.org/10.1137/19m1242069}.

\bibitem[Groth(2016)]{DBLPconfeurocryptGroth16}
Jens Groth.
\newblock On the size of pairing-based non-interactive arguments.
\newblock In Marc Fischlin and Jean{-}S{\'{e}}bastien Coron, editors, \emph{Advances in Cryptology - {EUROCRYPT} 2016 - 35th Annual International Conference on the Theory and Applications of Cryptographic Techniques, Vienna, Austria, May 8-12, 2016, Proceedings, Part {II}}, volume 9666 of \emph{Lecture Notes in Computer Science}, pages 305--326. Springer, 2016.
\newblock \doi{10.1007/978-3-662-49896-5\_11}.
\newblock URL \url{https://doi.org/10.1007/978-3-662-49896-5\_11}.

\bibitem[Gur et~al.(2024)Gur, Jahanara, Khodabandeh, Rajgopal, Salamatian, and Shinkar]{DBLPconfstocGurJKRSS24}
Tom Gur, Mohammad~Mahdi Jahanara, Mohammad~Mahdi Khodabandeh, Ninad Rajgopal, Bahar Salamatian, and Igor Shinkar.
\newblock On the power of interactive proofs for learning.
\newblock In Bojan Mohar, Igor Shinkar, and Ryan O'Donnell, editors, \emph{Proceedings of the 56th Annual {ACM} Symposium on Theory of Computing, {STOC} 2024, Vancouver, BC, Canada, June 24-28, 2024}, pages 1063--1070. {ACM}, 2024.
\newblock \doi{10.1145/3618260.3649784}.
\newblock URL \url{https://doi.org/10.1145/3618260.3649784}.

\bibitem[Hoeffding(1963)]{hoeffding1963probability}
Wassily Hoeffding.
\newblock Probability inequalities for sums of bounded random variables.
\newblock \emph{Journal of the American Statistical Association}, pages 13--30, 1963.
\newblock \doi{doi.org/10.2307/2282952}.
\newblock URL \url{https://doi.org/10.2307/2282952}.

\bibitem[Jun et~al.(2018)Jun, Li, Ma, and Zhu]{DBLPconfnipsJun0MZ18}
Kwang{-}Sung Jun, Lihong Li, Yuzhe Ma, and Xiaojin~(Jerry) Zhu.
\newblock Adversarial attacks on stochastic bandits.
\newblock In Samy Bengio, Hanna~M. Wallach, Hugo Larochelle, Kristen Grauman, Nicol{\`{o}} Cesa{-}Bianchi, and Roman Garnett, editors, \emph{Advances in Neural Information Processing Systems 31: Annual Conference on Neural Information Processing Systems 2018, NeurIPS 2018, December 3-8, 2018, Montr{\'{e}}al, Canada}, pages 3644--3653, 2018.
\newblock URL \url{https://proceedings.neurips.cc/paper/2018/hash/85f007f8c50dd25f5a45fca73cad64bd-Abstract.html}.

\bibitem[Karnin et~al.(2013)Karnin, Koren, and Somekh]{DBLPconficmlKarninKS13}
Zohar~Shay Karnin, Tomer Koren, and Oren Somekh.
\newblock Almost optimal exploration in multi-armed bandits.
\newblock In \emph{Proceedings of the 30th International Conference on Machine Learning, {ICML} 2013, Atlanta, GA, USA, 16-21 June 2013}, volume~28 of \emph{{JMLR} Workshop and Conference Proceedings}, pages 1238--1246. JMLR.org, 2013.
\newblock URL \url{http://proceedings.mlr.press/v28/karnin13.html}.

\bibitem[Kearns et~al.(2002)Kearns, Mansour, and Ng]{DBLPjournalsmlKearnsMN02}
Michael~J. Kearns, Yishay Mansour, and Andrew~Y. Ng.
\newblock A sparse sampling algorithm for near-optimal planning in large markov decision processes.
\newblock \emph{Mach. Learn.}, 49\penalty0 (2-3):\penalty0 193--208, 2002.
\newblock \doi{10.1023/A:1017932429737}.
\newblock URL \url{https://doi.org/10.1023/A:1017932429737}.

\bibitem[Lai and Robbins(1985)]{lai1985asymptotically}
Tze~Leung Lai and Herbert Robbins.
\newblock Asymptotically efficient adaptive allocation rules.
\newblock \emph{Advances in applied mathematics}, 6\penalty0 (1):\penalty0 4--22, 1985.

\bibitem[Lattimore and Szepesv{\'a}ri(2020)]{lattimore2020bandit}
Tor Lattimore and Csaba Szepesv{\'a}ri.
\newblock \emph{Bandit algorithms}.
\newblock Cambridge University Press, 2020.

\bibitem[Mannor and Tsitsiklis(2004)]{DBLPjournalsjmlrMannorT04}
Shie Mannor and John~N. Tsitsiklis.
\newblock The sample complexity of exploration in the multi-armed bandit problem.
\newblock \emph{J. Mach. Learn. Res.}, 5:\penalty0 623--648, 2004.
\newblock URL \url{https://jmlr.org/papers/volume5/mannor04b/mannor04b.pdf}.

\bibitem[Mutreja and Shafer(2023)]{DBLPconfcoltMutrejaS23}
Saachi Mutreja and Jonathan Shafer.
\newblock {PAC} verification of statistical algorithms.
\newblock In Gergely Neu and Lorenzo Rosasco, editors, \emph{The Thirty Sixth Annual Conference on Learning Theory, {COLT} 2023, 12-15 July 2023, Bangalore, India}, volume 195 of \emph{Proceedings of Machine Learning Research}, pages 5021--5043. {PMLR}, 2023.
\newblock URL \url{https://proceedings.mlr.press/v195/mutreja23a.html}.

\bibitem[Nisan et~al.(2007)Nisan, Roughgarden, Tardos, and Vazirani]{DBLPbookscuNRTV2007}
Noam Nisan, Tim Roughgarden, {\'{E}}va Tardos, and Vijay~V. Vazirani, editors.
\newblock \emph{Algorithmic Game Theory}.
\newblock Cambridge University Press, 2007.
\newblock ISBN 9780511800481.
\newblock \doi{10.1017/CBO9780511800481}.
\newblock URL \url{https://doi.org/10.1017/CBO9780511800481}.

\bibitem[Rubinstein(2017)]{DBLPjournalssigecomRubinstein17}
Aviad Rubinstein.
\newblock Settling the complexity of computing approximate two-player nash equilibria.
\newblock \emph{SIGecom Exch.}, 15\penalty0 (2):\penalty0 45--49, 2017.
\newblock \doi{10.1145/3055589.3055596}.
\newblock URL \url{https://doi.org/10.1145/3055589.3055596}.

\bibitem[Shamir(1992)]{DBLPjournalsjacmShamir92}
Adi Shamir.
\newblock {IP} = {PSPACE}.
\newblock \emph{J. {ACM}}, 39\penalty0 (4):\penalty0 869--877, 1992.
\newblock \doi{10.1145/146585.146609}.
\newblock URL \url{https://doi.org/10.1145/146585.146609}.

\bibitem[Sutton and Barto(2018)]{DBLPbookslibSuttonB2018}
Richard~S. Sutton and Andrew~G. Barto.
\newblock \emph{Reinforcement learning - an introduction, 2nd Edition}.
\newblock {MIT} Press, 2018.
\newblock URL \url{http://www.incompleteideas.net/book/the-book-2nd.html}.

\bibitem[Zhang et~al.(2022)Zhang, Chen, Zhu, and Sun]{DBLPconfaistatsZhangC0S22}
Xuezhou Zhang, Yiding Chen, Xiaojin Zhu, and Wen Sun.
\newblock Corruption-robust offline reinforcement learning.
\newblock In Gustau Camps{-}Valls, Francisco J.~R. Ruiz, and Isabel Valera, editors, \emph{International Conference on Artificial Intelligence and Statistics, {AISTATS} 2022, 28-30 March 2022, Virtual Event}, volume 151 of \emph{Proceedings of Machine Learning Research}, pages 5757--5773. {PMLR}, 2022.
\newblock URL \url{https://proceedings.mlr.press/v151/zhang22c.html}.

\end{thebibliography}
